\documentclass[letter]{aa}  

\usepackage{graphicx}
\usepackage{txfonts}
\begin{document}

   \title{The intervelocity of galaxy pairs in $\Lambda$CDM}

   \subtitle{The observed velocity peak at $\sim$130\,km\,s$^{-1}$ is not unique to MOND}

   \author{Marcel S. Pawlowski
          \inst{1}
  \and
   Kosuke Jamie Kanehisa
          \inst{1}
  \and
   Salvatore Taibi
          \inst{1}
  \and
   Pengfei Li
          \inst{1}\fnmsep\thanks{Humboldt Fellow}
          }

   \institute{Leibniz-Institute for Astrophysics,
              An der Sternwarte 16, 14482 Potsdam, Germany\\
              \email{mpawlowski@aip.de, kkanehisa@aip.de, staibi@aip.de, pli@aip.de}
             }

   \date{Received March 28, 2022; accepted July 16, 2022}

 
  \abstract
   {Observational studies of pairs of galaxies have uncovered that their differential line-of-sight velocities indicate the presence of a peak in their three-dimensional intervelocity distribution at $130-150\,\mathrm{km\,s}^{-1}$. It had been argued that galaxy pairs in the standard model of cosmology, $\Lambda$CDM, should not exhibit such an intervelocity peak, while Modified Newtonian Dynamics (MOND) predicts such a preferred intervelocity for paired galaxies. 
   However, no direct comparison with $\Lambda$CDM applying the same selection criteria and methodology as the observational studies has been performed yet, placing the comparison on unsure footing.}
   {To rectify this, we investigate this potential challenge for $\Lambda$CDM by determining whether an analog of the observed intervelocity peak is present in galaxy pairs within the IllustrisTNG-300 cosmological simulation.}
   {We identify galaxy pairs following the observational study’s selection criteria, measure their projected velocity difference, and analyse both the de-projected as well as the full velocity difference for this galaxy pair sample in the simulation.}
   {We recover a deprojected intervelocity peak at ${\sim}130\,\mathrm{km\,s}^{-1}$\ for galaxy pairs selected from the simulation. The full three-dimensional velocity information available for the pairs in the simulation also reveals a clear preference for this intervelocity.}
   {The intervelocity peak among galaxy pairs does not appear to be a feature unique to MOND, but is also present in $\Lambda$CDM. It can thus not be claimed as a unique success of either theory over the other. Developing the galaxy pair intervelocity into a test of gravity in the low acceleration regime will require more detailed studies to identify measurable differences in the models.}

   \maketitle
%

\section{Introduction}

 \citet{2018AstBu..73..310N} have obtained one of the largest samples of galaxy pairs with spectroscopic velocity information, using observed galaxies from the HyperLEDA database \citep{2014A&A...570A..13M}. The large sample of 13114 pairs allows to (statistically) deprojected the distribution in line-of-sight velocity difference ${\Delta}v$\ between pair members to obtain the distribution of full three-dimensional intervelocities $v_\mathrm{deproj}$\ \citep{2018A&A...614A..45N}. Applying this deprojection to their pair catalog, \citet{2020A&A...641A.115N} have identified a robust peak at $v_\mathrm{deproj} \sim 150\,\mathrm{km\,s}^{-1}$. \citet{2022MNRAS.510.2167S, 2022arXiv220213766S} confirmed this intervelocity peak with an expanded input galaxy catalog, reporting its location to be at $v_\mathrm{deproj} = 132 \pm 5\,\mathrm{km\,s}^{-1}$.

\citet{2022MNRAS.510.2167S} proposed that galaxy pairs can be used to test gravity in the low acceleration regime, specifically to investigate the dark matter alternative Modified Newtonian Dynamics (MOND, \citealt{1983ApJ...270..365M, 2012LRR....15...10F}). Since the MOND force at low accelerations ($a{\ll}a_0$, as realized for galaxy pairs) scales as $1/r$, the total mutual velocity of a pair of galaxies in circular orbit is set by their mass and independent of separation $r$. The absolute magnitude limit of the observed pairs ($M_\mathrm{B}<-18.5$) sets a typical galaxy mass scale, which then translates into a typical intervelocity. \citet{2022arXiv220213766S} demonstrate that a realistic B-band mass-to-light ratio of $M/L{\sim}1$ reproduces the observed peak position in MOND.

In contrast, \citet{2022MNRAS.510.2167S} argued that the velocity peak is ``hard to justify in the context of numerical simulations of cosmological structure formation''. This assessment appears mostly based on \citet{2013MNRAS.436.1765M}, who study galaxy pairs in the Millennium simulation \citep{2005Natur.435..629S}. They find a broad distribution in intervelocities of pairs identified by three-dimensional separation. Such a pair-selection is possible in simulations where full phase-space information is available, but not directly comparable to the observational pair-finding algorithm that must rely on only projected positions and redshift velocities. The simulation study also allows much more extreme mass ratios between paired galaxies, while the observational study applies a stricter isolation criterion. Nevertheless, pairs in \citet{2013MNRAS.436.1765M} do display an intervelocity peak at $\sim200\,\mathrm{km\,s}^{-1}$\ if restricted to a central-satellite subsample. Considering only pairs whose mutual binding energy dominates, as opposed to them being bound to a third object, their intervelocity distribution even reaches a maximum between $100~\mathrm{and}~200\,\mathrm{km\,s}^{-1}$. Therefore, it is not firmly established that the observed peak is absent in a $\Lambda$CDM simulation, if galaxy pairs were selected following the observational study's approach.

We set out to determine whether an intervelocity peak between paired galaxies is indeed absent in a $\Lambda$CDM context. Were this the case, the observed peak would constitute both a serious challenge to $\Lambda$CDM and a success for MOND.
We approach this task by identifying galaxy pairs in the modern hydrodynamical cosmological simulation IllustrisTNG-300 \citep{2019ComAC...6....2N, 2018MNRAS.475..676S, 2018MNRAS.475..624N, 2018MNRAS.475..648P}. To ensure direct comparability, we mock-observe the simulation to only consider projected positions and line-of-sight velocities when identifying pairs, follow the pair-finding algorithm of the observational studies, and also apply the same statistical deprojection algorithm to obtain the three-dimensional intervelocities.


\section{Simulation, Galaxy and Pair Selection}

\citet{2018AstBu..73..310N}'s galaxy sample covers those with B-band magnitudes $M_B < -18.5$ and heliocentric redshifts $2500 < v < 16500 \,\mathrm{km\,s}^{-1}$, which -- in the absence of peculiar velocities -- corresponds to distances of $37 < d < 244 \,\mathrm{Mpc}$, adopting a Hubble constant of $H_0 = 67.74\,\mathrm{km\,s}^{-1}\,\mathrm{Mpc}^{-1}$. To make a fair selection of simulated galaxies using the same constraints, we require a hydrodynamic simulation with a box size exceeding $200$\,Mpc.
We elect to use IllustrisTNG-300, a hydrodynamic cosmological simulation spanning a periodic box of length $L_{\mathrm{box}} = 300 \,\mathrm{Mpc}$ under a \citet{2016A&A...594A..13P} cosmology.

Following \citet{2018AstBu..73..310N}, we create a catalog of simulated subhalo-based galaxies with $M_B < -18.5$, and mock-observed along each of the simulation's three orthonormal axes. For axis $a \in \{ x, y, z \}$, we place an observer on the $a = 0$ plane at the edge of the simulation box and observe along the positive $a$ axis -- parallax is assumed to be negligible due to the large distances involved. We calculate the redshift of each galaxy at distance $D_a$ along an axis $a$ as
\begin{equation}
    v = H_0 \, d_a + v_{\mathrm{pec}},
\end{equation}
where $v_{\mathrm{pec}}$ is the peculiar velocity of the galaxy in the simulation's co-moving frame. By rejecting galaxies with a redshift outside $2500 < v < 16500 \,\mathrm{km\,s}^{-1}$, we obtain samples between $109000 - 116000$ individual galaxies depending on the axis\footnote{We adopt this redshift range for comparability to the observational study, but have tested that a broader range, while increasing the sample size, does not change our results.}. Limitations of this mock-observational approach are discussed in Appendix \ref{appendix:othertests}. In the following, we focus on the $z$-axis projection to avoid cluttered figures, but comparisons between the different projections are shown in Appendix \ref{Appendix:DifferentDirections}.

We investigate a \textit{base} sample directly using the positions and velocities of the simulation's galaxy catalog, without introducing any errors. We also generate a \textit{full} sample with an apparent magnitude cut-off at $m_\mathrm{B} = 19$\ similar to the observational galaxy catalog, corresponding to an apparent magnitude of $M_{\mathrm{B}} = -18$ observed at the maximum redshift of $16000 \,\mathrm{km\,s}^{-1}$\ (see Fig. 1 in \citealt{2018AstBu..73..310N}). For each simulated galaxy, velocity errors are drawn with replacement from the errors in the magnitude-limited HyperLEDA galaxy database to ensure that our overall error distribution follows that of the observational catalog. The velocity $v$ of the galaxy is then displaced by a random draw from a Gaussian distribution with the error as its standard deviation. We also generate two randomized samples: \textit{shuffled velocities} for which the galaxy velocities are re-assigned randomly while their positions (and thus spatial clustering) are kept, and \textit{random positions} with positions randomly selected from a uniform distribution within the simulation volume.

To obtain the list of galaxy pair candidates from our mock galaxy catalogs, we closely follow the successive selection criteria of the observational study \citep{2018AstBu..73..310N}. Specifically, we identify pairs for which the two galaxies  
\begin{itemize}
    \item have B-band magnitudes $M_\mathrm{B} < -18.5$
    \item have a projected line-of-sight velocity difference of $\Delta v = \Delta v_\mathrm{z} < 500\,\mathrm{km\,s}^{-1}$
    \item have a projected separation of $r_\mathrm{p} = \sqrt{(\Delta x)^2 + (\Delta y)^2} < 1\,\mathrm{Mpc}$
    \item fulfill a reciprocity criterion to exclude multiples or hierarchical groups.
\end{itemize}
We then require these candidate pairs to be isolated using the ratio $\rho = r_3 / r_\mathrm{p}$, where $r_3$\ is the separation of the next-nearest galaxy (within $500\,\mathrm{km\,s}^{-1}$\ from the pair) from the pair midpoint. If no nearby galaxy with $M_\mathrm{B} < -18.5$\ is found, the magnitude range is expanded to include galaxies up to 2.5 magnitudes fainter than the fainter pair member. Only galaxies with $\rho > 2.5$\ are retained, resulting in 7840 (8595) pairs for our base (full) galaxy catalog. Stricter limits on $\rho$\ can be applied to look at particularly isolated galaxy pair samples.

\section{Comparison to Observed Galaxy Pair Catalog}

   \begin{figure*}
   \centering
   \includegraphics[width=0.3\hsize]{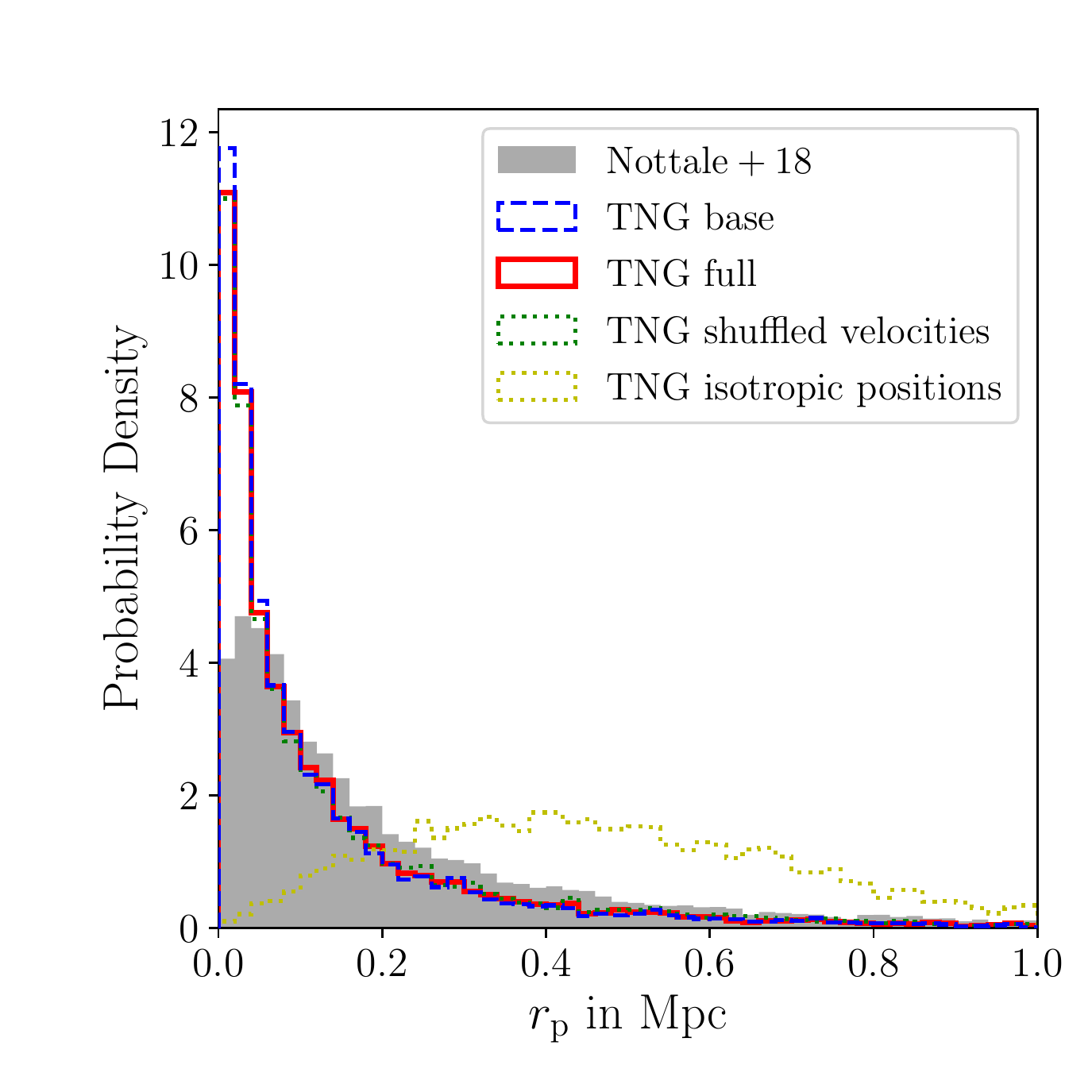}
   \includegraphics[width=0.3\hsize]{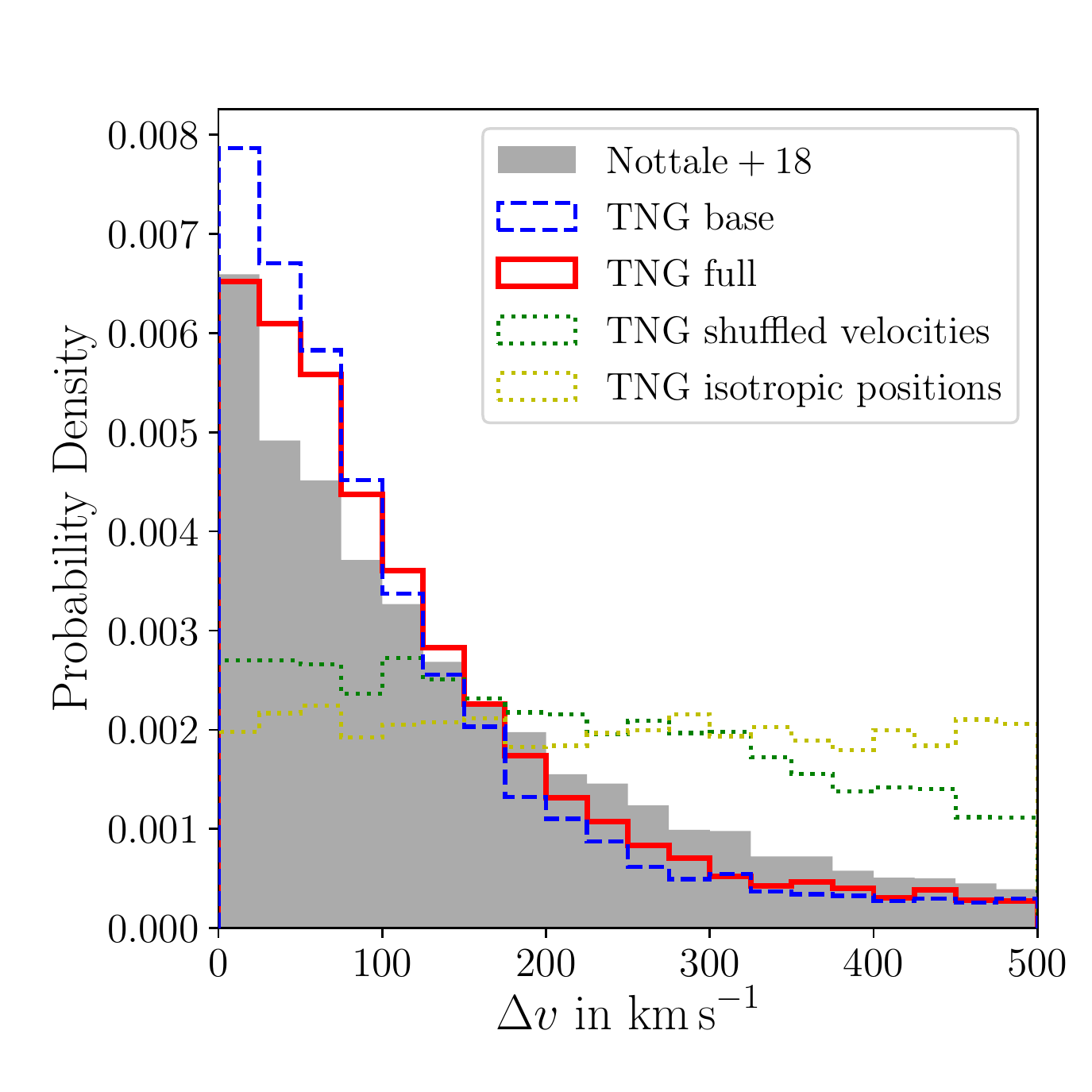}
   \includegraphics[width=0.3\hsize]{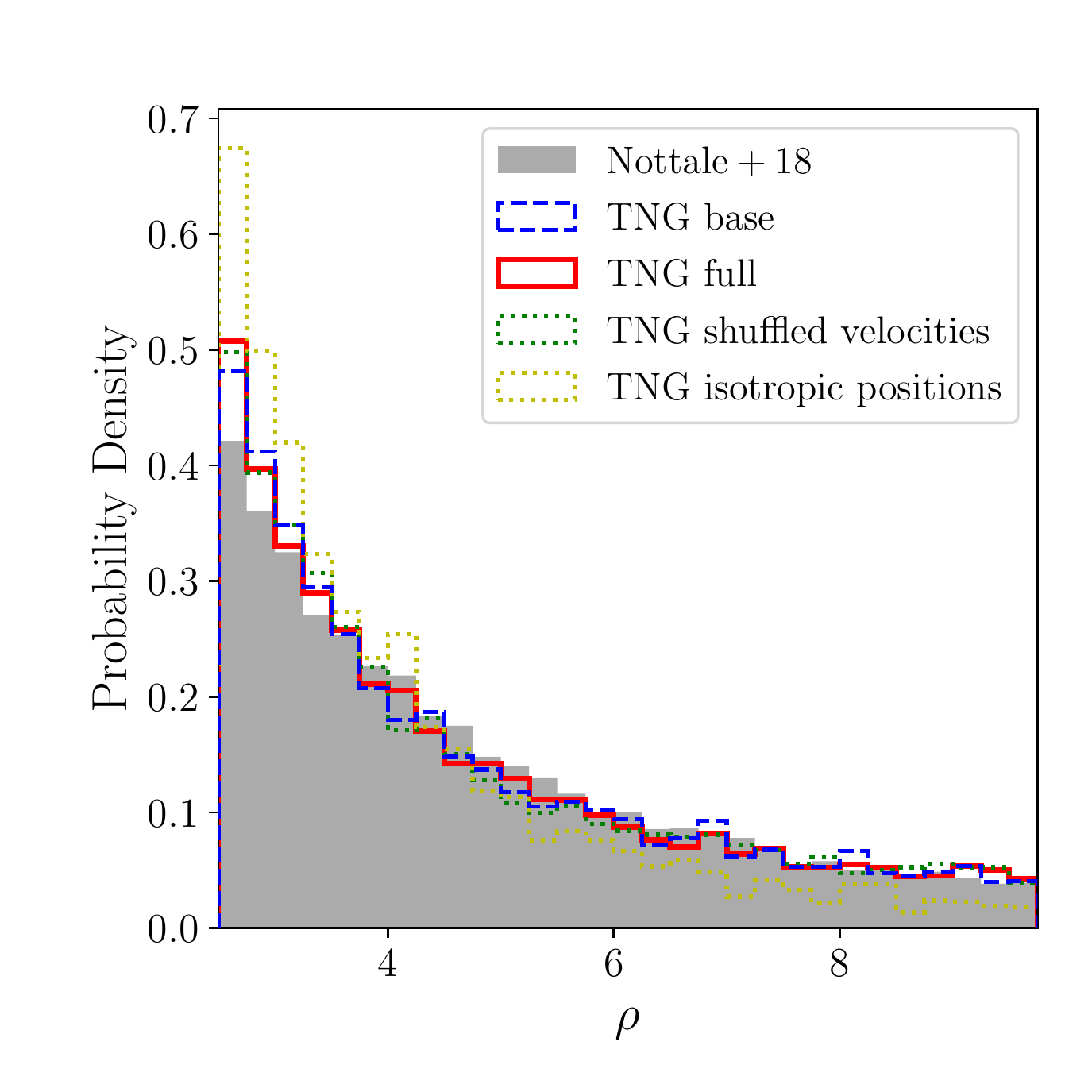}
      \caption{Distributions of galaxy-galaxy pair properties. The left panel shows the projected separation $r_\mathrm{p}$, the middle panel shows the difference in line-of-sight velocity $\Delta v$, and the right panel shows the isolation criterion $\rho$). In all panels, the observed galaxy pairs are highlighted as a grey filled histogram, while the mock-observed simulation data is shown as lines for the base (blue dashed) and full (red solid) sample, as well as the two randomized test samples with shuffled velocities but retained positions (green dotted) and randomized positions (yellow dotted).
              }
         \label{FigPairParameters}
   \end{figure*}

In Fig. \ref{FigPairParameters} we compare the observed galaxy pairs \citep{2018AstBu..73..310N} with our base and full galaxy pair samples, and the two randomized samples.
Observed and simulated galaxy pairs follow comparable distributions in the on-sky separation, with most having $r_\mathrm{p} < 200\,\mathrm{kpc}$. One difference is the much reduced peak at small $r_\mathrm{p}$\ for the observed sample: very small separations are over-represented in the simulated pairs. This is likely caused by an observational bias. In the simulation, all galaxy positions and velocities are known and included in the input catalogs. In contrast, in the observational galaxy catalog galaxies that are projected on top of each other, or have only very small separations, are likely either not included, are included as one object, or do not both have measured redshift velocities due to fiber positioning constraints in surveys. To address this difference, we have confirmed that rejecting galaxy pairs with $r_\mathrm{p} < 20\,\mathrm{kpc}$\ in our simulation's pair catalog does not affect our later results. Randomizing only the galaxy velocities retains their spatial clustering and thus distribution in $r_\mathrm{p}$, while randomized positions result in a much broader distribution.

The middle panel of Fig. \ref{FigPairParameters} shows the difference in the line-of-sight velocities $\Delta v$\ of the galaxies in a pair, the quantity to be deprojected to obtain the distribution of three-dimensional intervelocities of paired galaxies. The observed and the simulated galaxy pairs again show a similar behaviour, with an initially steeply declining distribution that flattens at higher $\Delta v$. This is slightly more pronounced even for the simulated full sample including velocity errors than for the observational sample, possibly related to underestimated errors or unaccounted systematics in the spectroscopic velocities, which bias towards larger velocity differences. Both randomized samples show much flatter distributions in $\Delta v$, suggesting that the shape of the distribution has a physical origin.

The right panel of Fig. \ref{FigPairParameters} shows the distribution of the isolation criterion $\rho$. A very similar decrease in the relative number of more isolated pairs is present in all samples, though pairs in the simulation show a slight excess at small $\rho$\ over the observational sample. This is likely a result of the simulation having a full coverage of all galaxies, compared to some degree of incompleteness in the observational catalog.

\citet{2018AstBu..73..310N} found that about 10\% of the galaxies with $M_\mathrm{B} < -18.5$\ are members of fairly isolated pairs ($\rho > 5$), and that highly isolated pairs ($\rho\geq10$) contribute 32.5\% of the total number of pairs. We find 4439 fairly isolated pairs in the base sample (8.1\% of 109603 galaxies) and 4892 in the full sample (8.9\% of 109609 galaxies), and that 34.0\% (34.5\%) of the pairs in the base (full) sample have $\rho\geq10$. The similarity between the simulated and observed pairs in these quantities gives confidence in our mock-observational approach, and the agreement also in the distributions discussed above might already be seen as a success of the simulation in reproducing a galaxy sample comparable to the observational study.

\section{Intervelocity Distribution of Galaxy Pairs}
\label{Sect:Intervelocity}

   \begin{figure*}
   \centering
   \includegraphics[width=0.3\hsize]{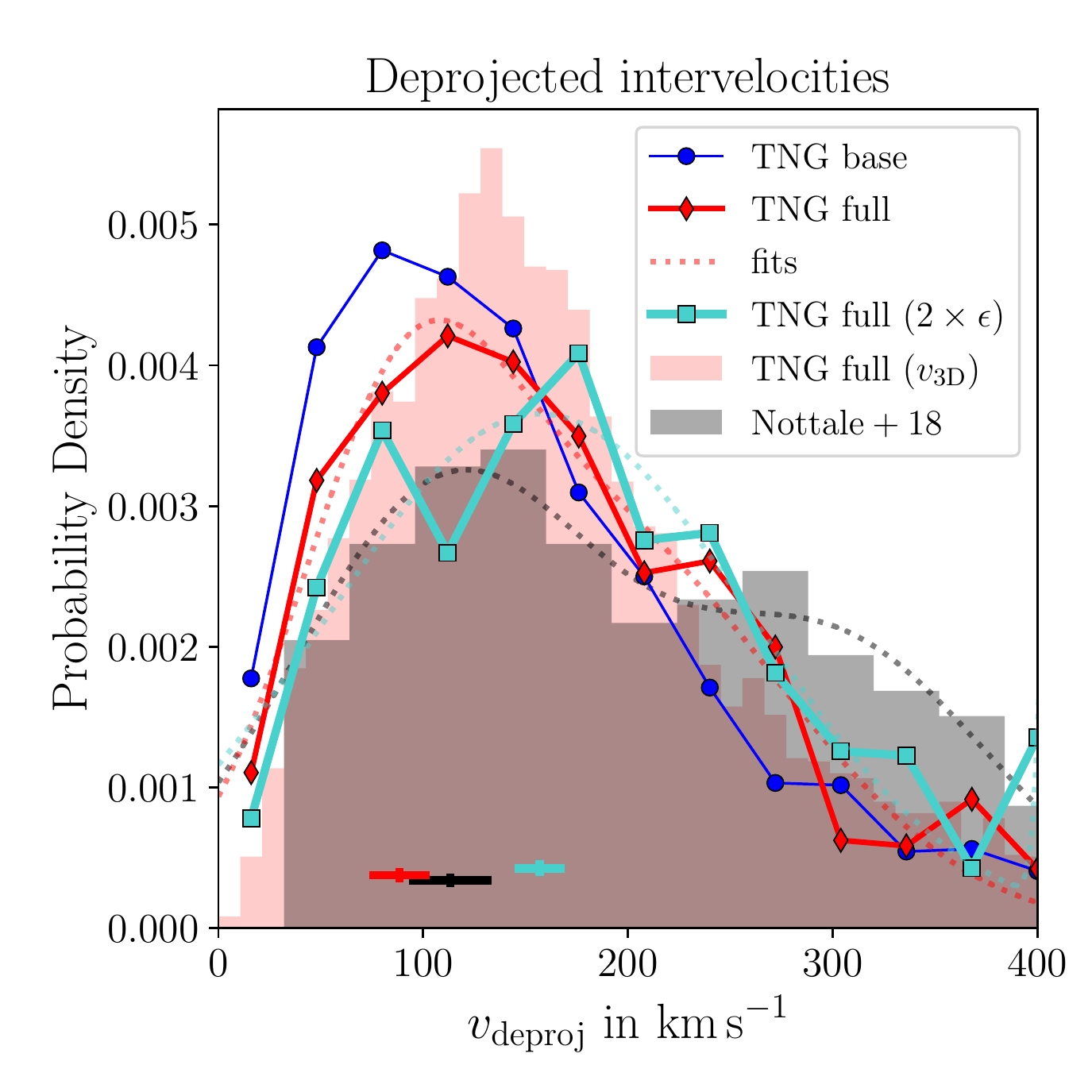}
   \includegraphics[width=0.3\hsize]{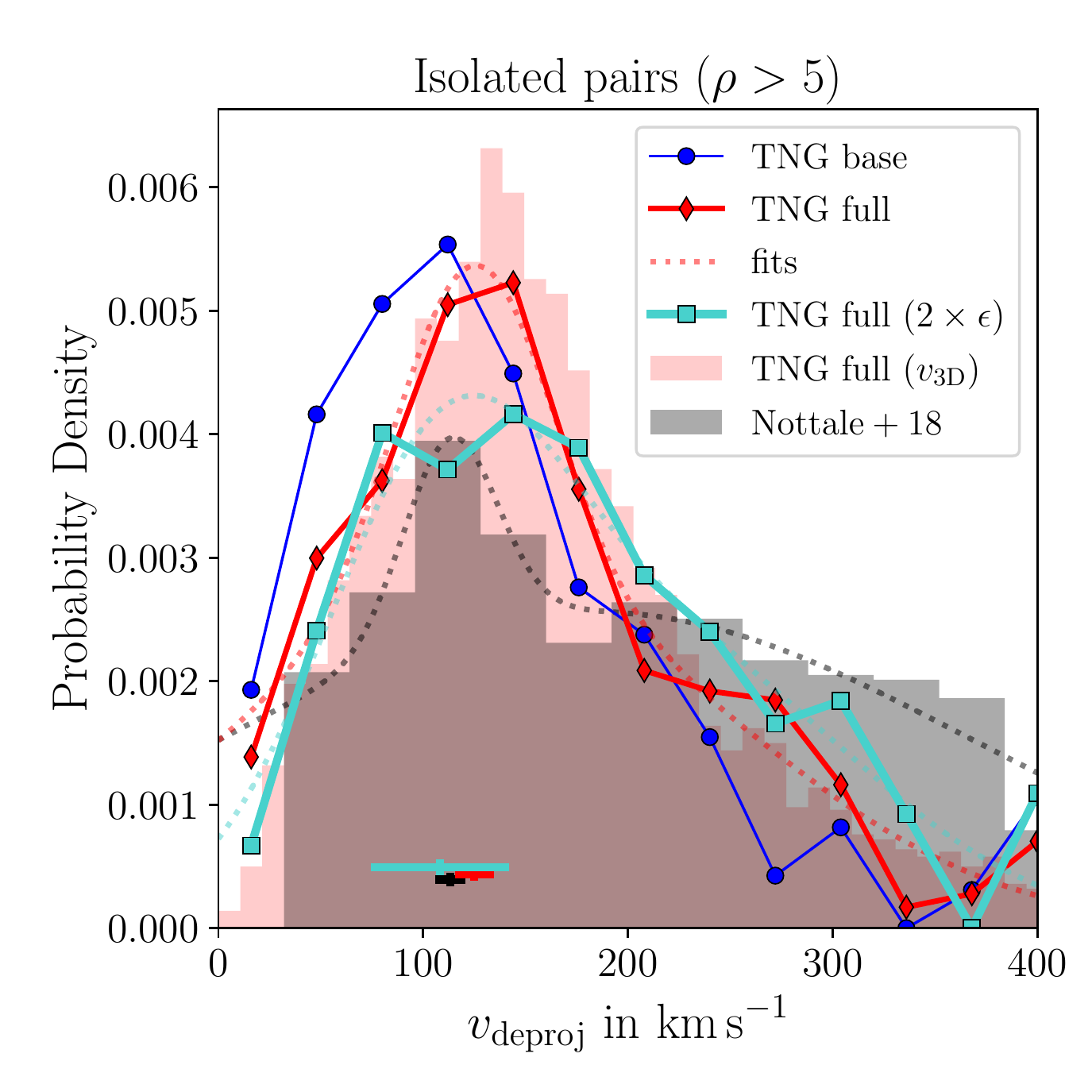}
   \includegraphics[width=0.3\hsize]{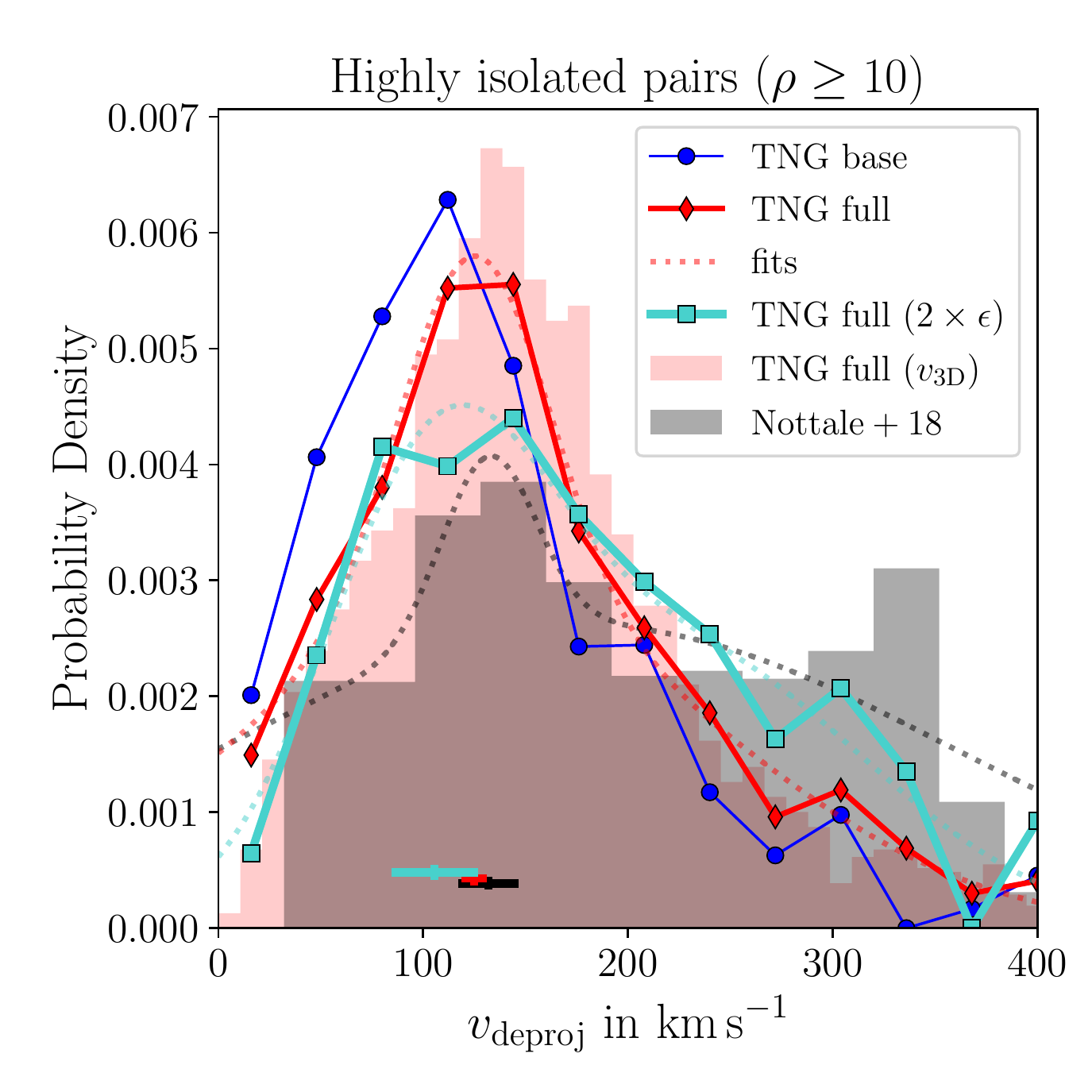}
      \caption{Deprojected velocities for the observed pairs (grey) and mock-observed simulation without (blue), with (red) and with doubled (turquoise) velocity errors. 
      The light-red histogram in the background shows the $v_\mathrm{3D}$\ intervelocity distribution for the full sample, as determined from the peculiar velocity difference of the galaxy pairs using the available full phase-space information of the simulation. Also shown are the two-Gaussian model fits to the histograms used to determine the positions of the velocity peak. The peak position and its $1\sigma$-uncertainty are indicated by solid lines at the bottom of the figure.
              }
         \label{FigDeprojections}
   \end{figure*}

   \begin{figure*}
   \centering
   \includegraphics[width=0.3\hsize]{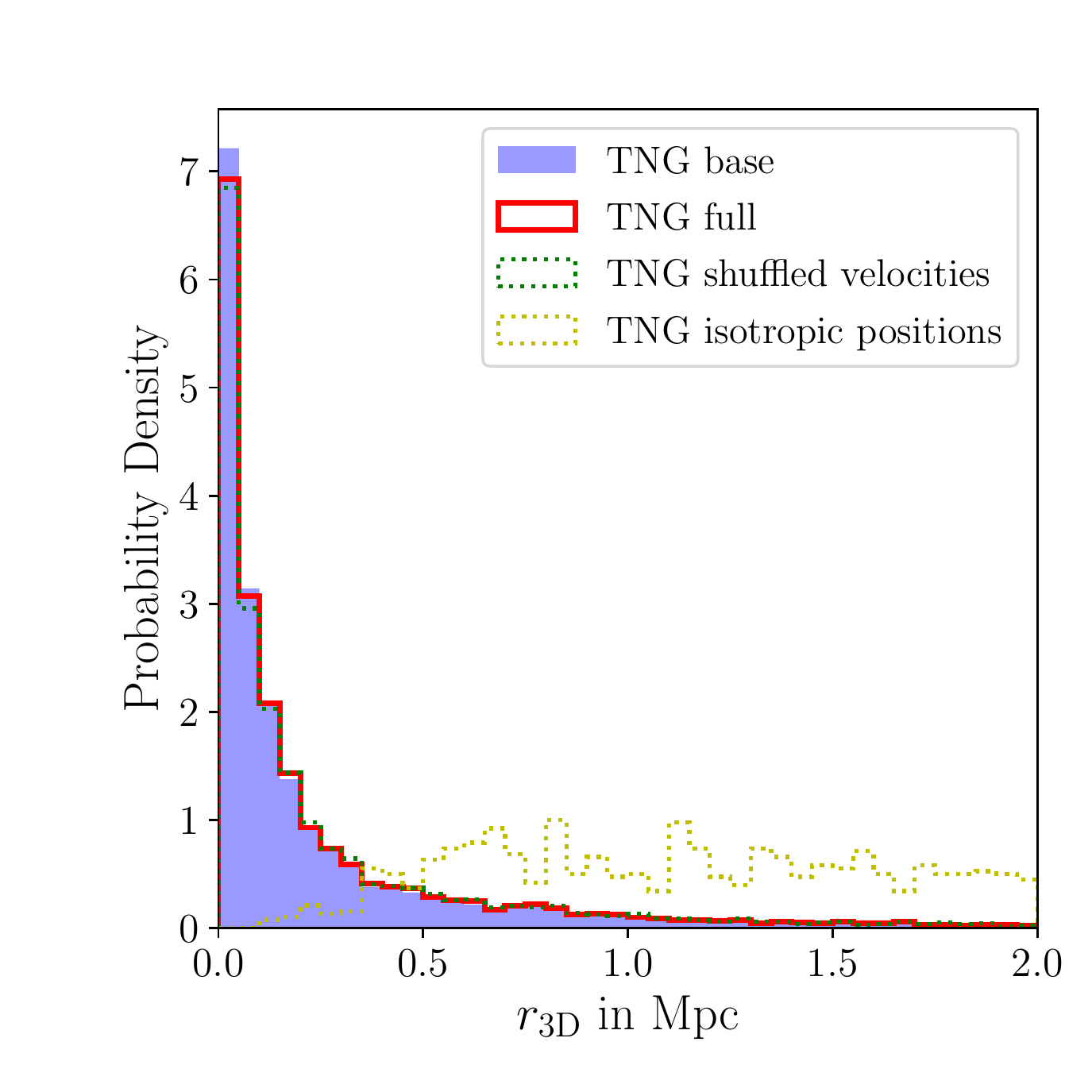}
   \includegraphics[width=0.3\hsize]{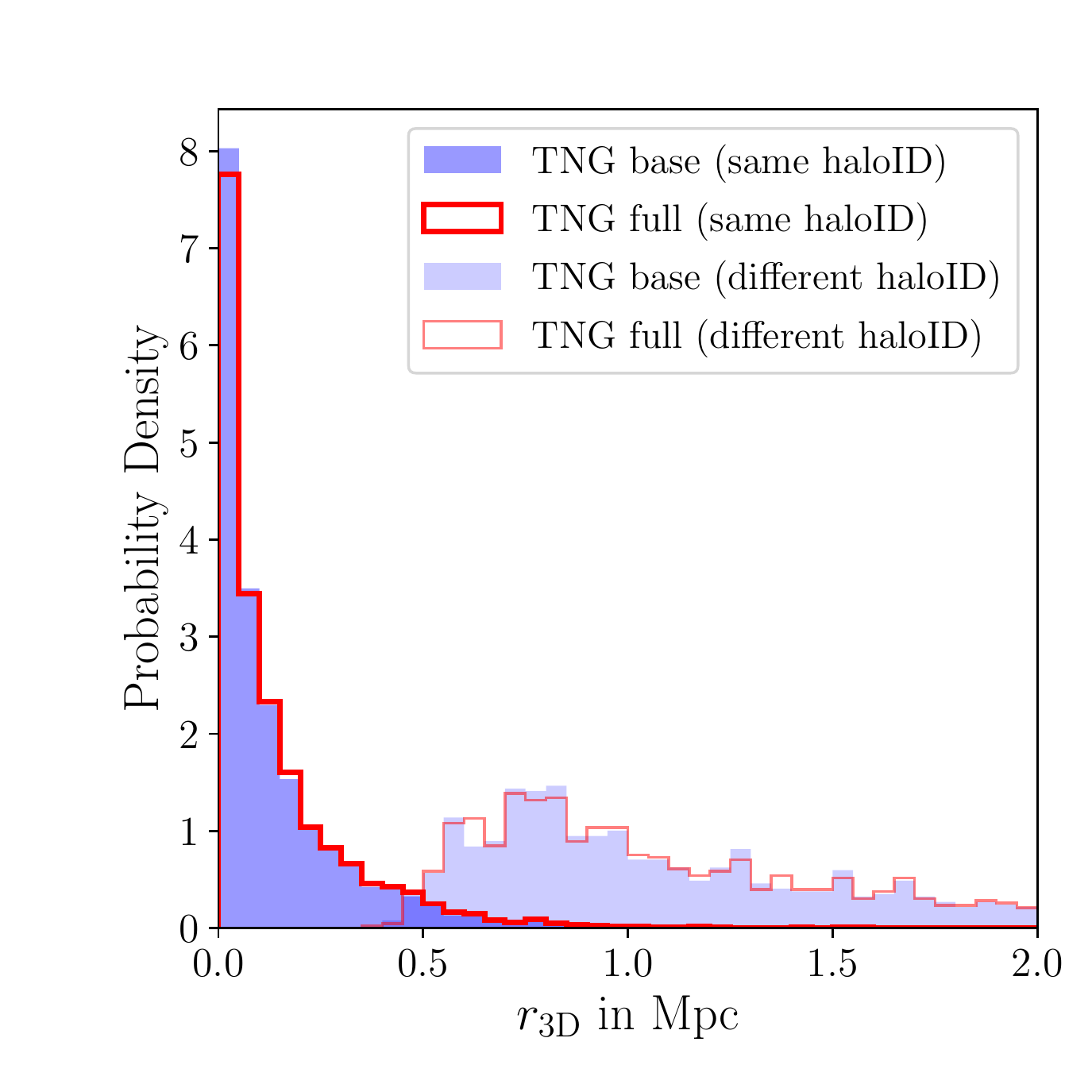}
   \includegraphics[width=0.3\hsize]{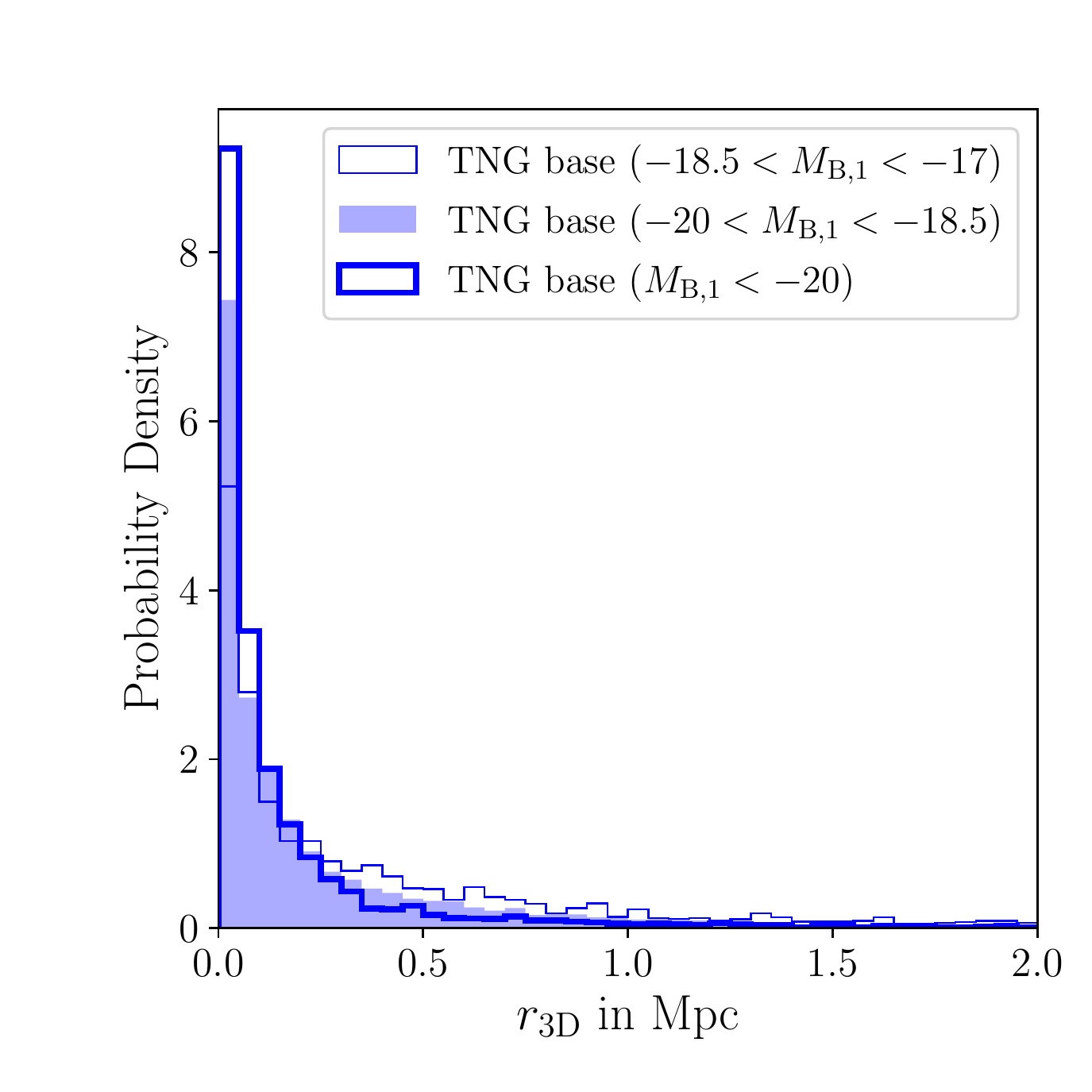}
   
   \includegraphics[width=0.3\hsize]{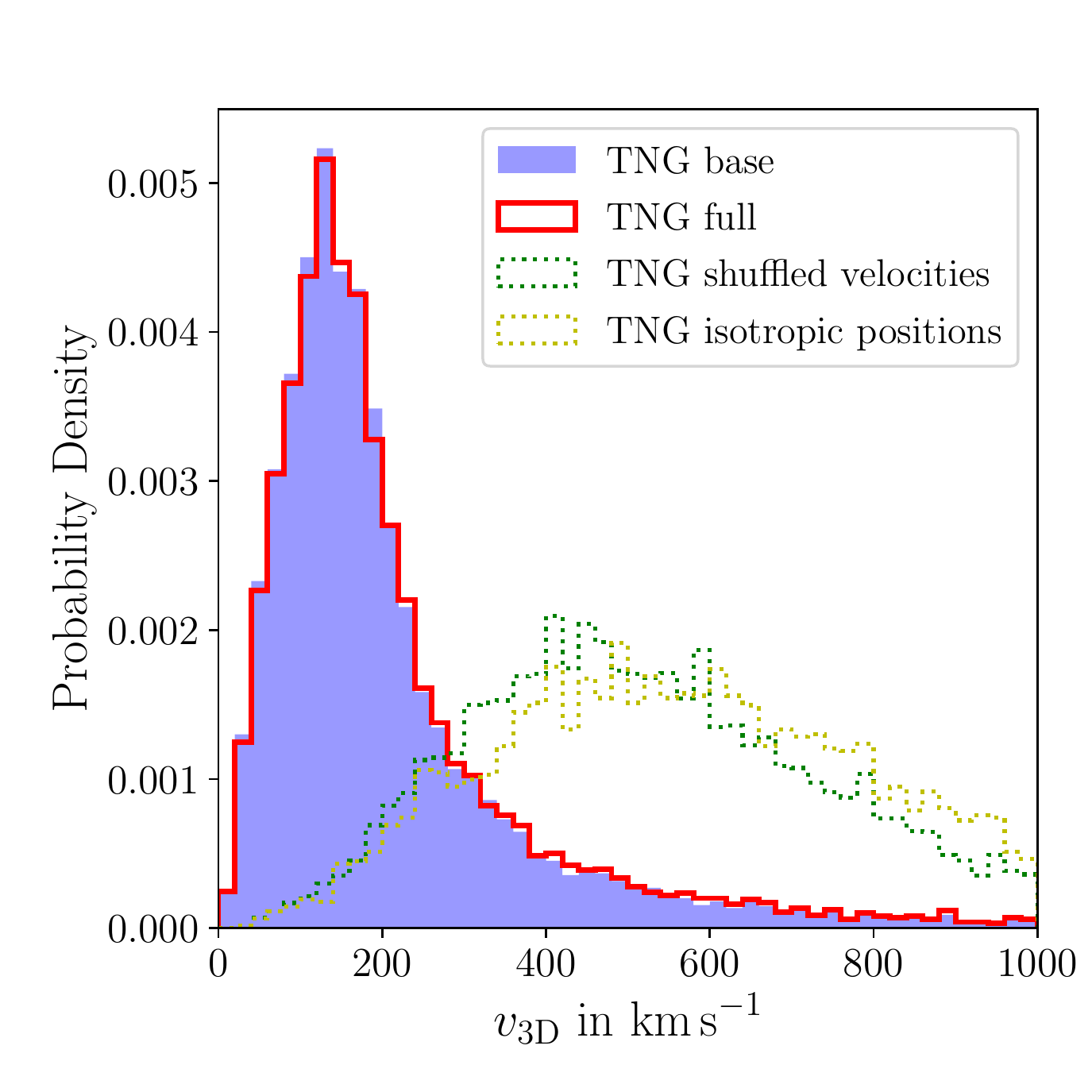}
   \includegraphics[width=0.3\hsize]{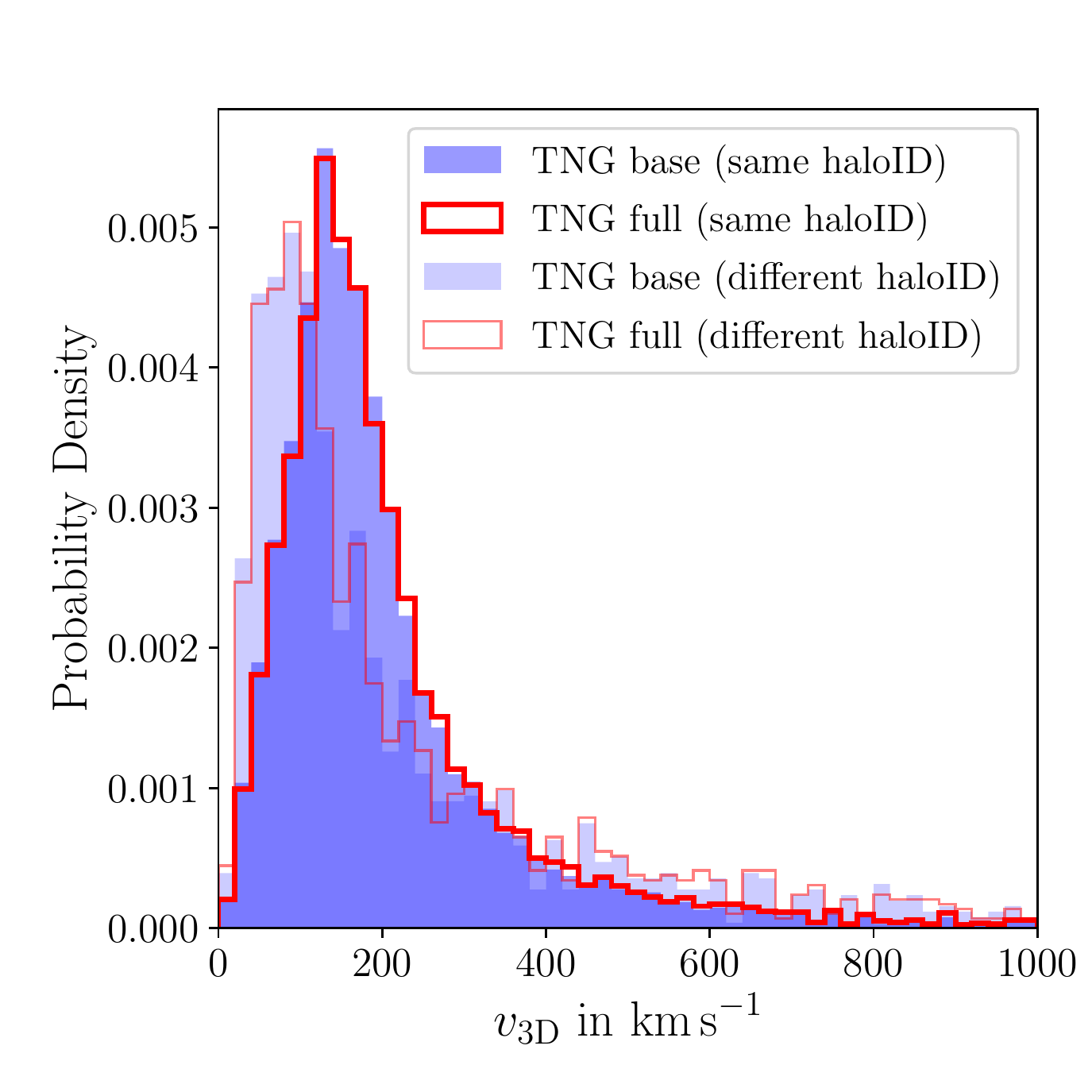}
   \includegraphics[width=0.3\hsize]{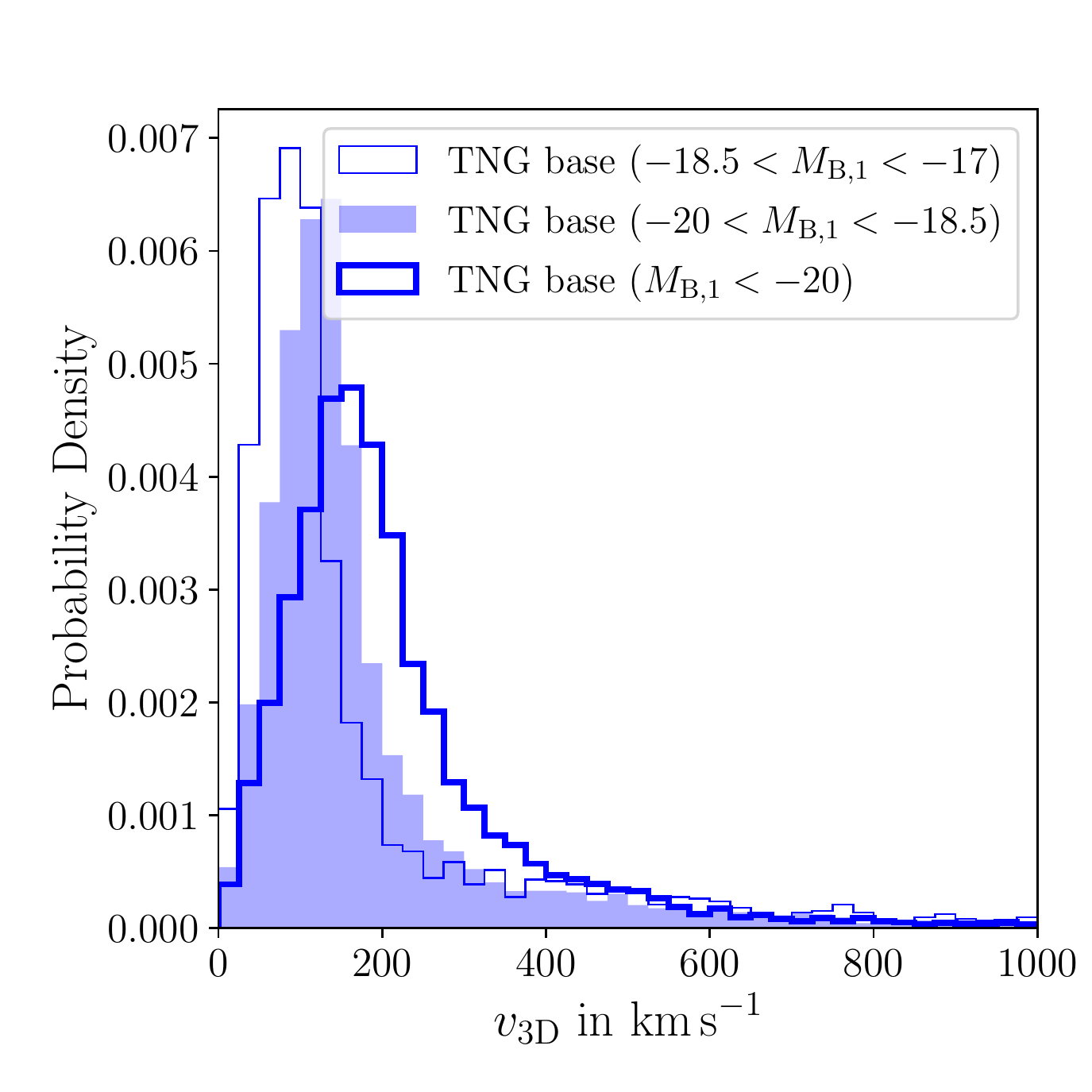}
      \caption{Full three-dimensional separation $r_\mathrm{3D}$\ ({\it top panels}) and 3D intervelocity $v_\mathrm{3D}$\ ({\it bottom panels}) for our base (blue shaded histograms) and full (red solid line) simulation galaxy pair catalogs. 
      The {\it left panels} compare with the two randomized samples for which either only the galaxy velocities were shuffled among all considered galaxies, or their positions were randomly drawn from a uniform distribution. The {\it middle panels} compare galaxy pairs which share a common halo ID in the simulation with those that do not, while the {\it right panels} compare results for different ranges of the pair primary's absolute magnitude.
              }
         \label{Fig3Dquantities}
   \end{figure*}

For best comparability with the observational studies, we initially focus on the de-projection of the velocity difference of our galaxy pair catalog into the 3D intervelocity distribution. The de-projection is made with the algorithm proposed by \citet{2018A&A...614A..45N}:
\begin{equation}
    P_v(v) = -v\Big[\frac{{\rm d}P_{v_z}(v_z)}{{\rm d}v_z}\Big]_v.
\end{equation}
The formula is based on the fact that a randomly oriented velocity $\vec{v}$ with a given value is equally projected along a given axis, i.e. the probability distribution of $v_z$ is constant from 0 to $|\vec{v}|$. The de-projected velocity distribution can then be derived by inverting the probability distribution of $v_z$. The derivative in equation (2) is calculated using the differences between constant intervals separated by two bins. This method proves more accurate and generates a smoother de-projected distribution \citep{2018A&A...614A..45N}. We binned the probability distribution of $\Delta v$ with a width of 32\,km\,s$^{-1}$\ to ensure a monotonically decreasing distribution. Other bin widths result in very similar plots, in particular the peaks at $\sim 130\,\mathrm{km\,s}^{-1}$\ are reliably recovered.

The similarity in $\Delta v$\ between the observed and simulated galaxy pairs (Fig. \ref{FigPairParameters}) suggests that their deprojected intervelocities might also be compatible. This is confirmed in Figure \ref{FigDeprojections}, which shows the results of the velocity deprojection for three choices of isolation. Both the observed and the simulated pair samples display clear intervelocity peaks at $v_\mathrm{deproj} \sim 130\,\mathrm{km\,s}^{-1}$.

To measure the positions of the prominent intervelocity peaks, we fit them with a double-Gaussian model to account for the peak and a broader background distribution. The peak positions are $v_\mathrm{deproj} = 88\pm13$, $125\pm7$, and $125\pm4,\mathrm{km\,s}^{-1}$\ for the simulated full sample ($\rho > 2.5$), fairly isolated pairs ($\rho > 5$), and highly isolated pairs (($\rho \geq 10$), respectively. For the observed pair catalog the same cuts in $\rho$\ result in peak positions of $v_\mathrm{deproj} = 113\pm18$, $113\pm5$, and $132\pm13,\mathrm{km\,s}^{-1}$, respectively. In all three cases, the peak positions agree within their $1\sigma$\ uncertainties between the simulated and the observed galaxy pairs. These intervelocity peaks are approximately the same intervelocity of $v_\mathrm{deproj}{\sim}150\,\mathrm{km\,s}^{-1}$\ reported by \citet{2020A&A...641A.115N} and $v_\mathrm{deproj}=132\pm5\,\mathrm{km\,s}^{-1}$\ reported by \citet{2022arXiv220213766S}. 

The peak in our base and full samples are more pronounced than the observed one. The reason might be that the simulation data is, by definition, much cleaner, while observational biases and underestimated or unaccounted-for measurement errors and systematics in the observational data set can wash out correlations, reducing the velocity peak strength, and add contamination by spurious pairs at higher intervelocities. We test this hypothesis with a simulated galaxy catalog for which we doubled velocity errors relative to the full sample (turquoise lines in Fig. \ref{FigDeprojections}). The resulting distribution of $v_\mathrm{deproj}$\ is indeed more consistent with the observed sample, the peak strength is reduced, but its position is not substantially affected.

The simulation data provides access to the full 6D phase-space of the galaxy pairs, such that we can also immediately investigate the pairs' 3D intervelocity $v_\mathrm{3D}$\ without deprojection. This is shown as the red filled histogram in Fig. \ref{FigDeprojections} for the full sample (and in Fig. \ref{FigDeprojections} for the base sample and various test samples). This confirms the de-projection analysis: for galaxy pairs identified in the simulation a clear intervelocity peak is present at $v_\mathrm{3D} = 132\pm1\,\mathrm{km\,s}^{-1}$, with a width of $\sigma_{v_\mathrm{3D}} = 61\pm2\,\mathrm{km\,s}^{-1}$.

\section{The Intervelocity Peak as a Physical Feature}
\label{subsect:tests}

To investigate the origin of the intervelocity peak in the simulation, we performed a number of tests and additional comparisons. These are illustrated in Fig. \ref{Fig3Dquantities}, showing distributions in the three-dimensional separation of the two pair members $r_\mathrm{3D}$\ and their full three-dimensional intervelocity $v_\mathrm{3D}$.

We first compared the simulated samples to the two randomized samples (left panels in Fig. \ref{Fig3Dquantities}). While the shuffled velocities sample preserves the spatial clustering and results in a very similar distribution in $r_\mathrm{3D}$\ as for the mock-observed samples, neither randomized sample reproduces the velocity distribution or its narrow peak. Instead, they display broad maxima at $v_\mathrm{3D} = 426\pm8$\ and $491\pm11\,\mathrm{km\,s}^{-1}$, respectively. This is close to the velocity limit of the pair selection algorithm and thus likely driven by that constraint. This suggests that the velocity peak identified in the simulation is not merely an emergent feature of the typical galaxy distribution or their typical peculiar velocities, but rather has a physical origin related to galaxies associated with each other.

To investigate such a physical association, we have determined which galaxy pairs identified in the mock-observed simulation share the same halo ID. These have been identified as being physically associated by the halo finder. We find that 83.4\% (82.5\%) of the galaxy pairs in the base (full) sample share a common halo ID, indicating that the galaxy pair finding criteria are very good at identifying truly associated galaxies, with only one sixth being contamination by spurious pairs. We also find that 75.2\% (74.4\%) of the pairs share a {\it unique} halo ID not shared by any other pair, indicating that only a small fraction of pairs are associated with massive halos (likely galaxy clusters) hosting more than one pair.

In the middle panels of Fig. \ref{Fig3Dquantities} we show distributions in $r_\mathrm{3D}$\ and $v_\mathrm{3D}$ for pairs with and without a common halo ID. Only galaxy pairs in a common halo have small $r_\mathrm{3D}$, as expected for physically associated systems. They also drive the bulk of the velocity peak at $v_\mathrm{3D} \sim 130\,\mathrm{km\,s}^{-1}$, while the (spurious) pairs belonging to different halos preferentially show lower intervelocities. This appears in line with their larger distance, resulting in lower mutual acceleration. Again fitting a two-Gaussian model, we find that pairs associated to the same halo display a peak position of $v_\mathrm{3D} = 138\pm1\,\mathrm{km\,s}^{-1}$\ for the full sample, while those belonging to different halos have a peak position at $v_\mathrm{3D} = 88\pm2\,\mathrm{km\,s}^{-1}$. The corresponding positions for the base sample do not differ substantially.

Finally, we compare different ranges in the absolute magnitude $M_\mathrm{B,1}$\ of the more luminous galaxy of each pair. This is based on an extended galaxy pair catalog where also fainter galaxies with $M_\mathrm{B,1} \leq -17$\ were considered. 
As shown in the right panels of Fig. \ref{Fig3Dquantities}, their spatial distributions are comparable, but pairs with more luminous primaries display higher intervelocities. We find peaks at $v_\mathrm{3D} = 86\pm1$, $117\pm1$, and $153\pm1,\mathrm{km\,s}^{-1}$\ for the subsequently more luminous samples in Fig. \ref{Fig3Dquantities}.
This is in line with a physical origin of the intervelocity peak as being caused by the mutual acceleration of paired galaxies, with more luminous (and thus on average more massive) primary galaxies resulting in higher accelerations, causing larger velocities. The absolute magnitude limit of the observational study then determines which galaxy masses contribute to the identified pairs most, implicitly setting the intervelocity peak of the sample.

   \begin{figure}
   \centering
   \includegraphics[width=0.75\hsize]{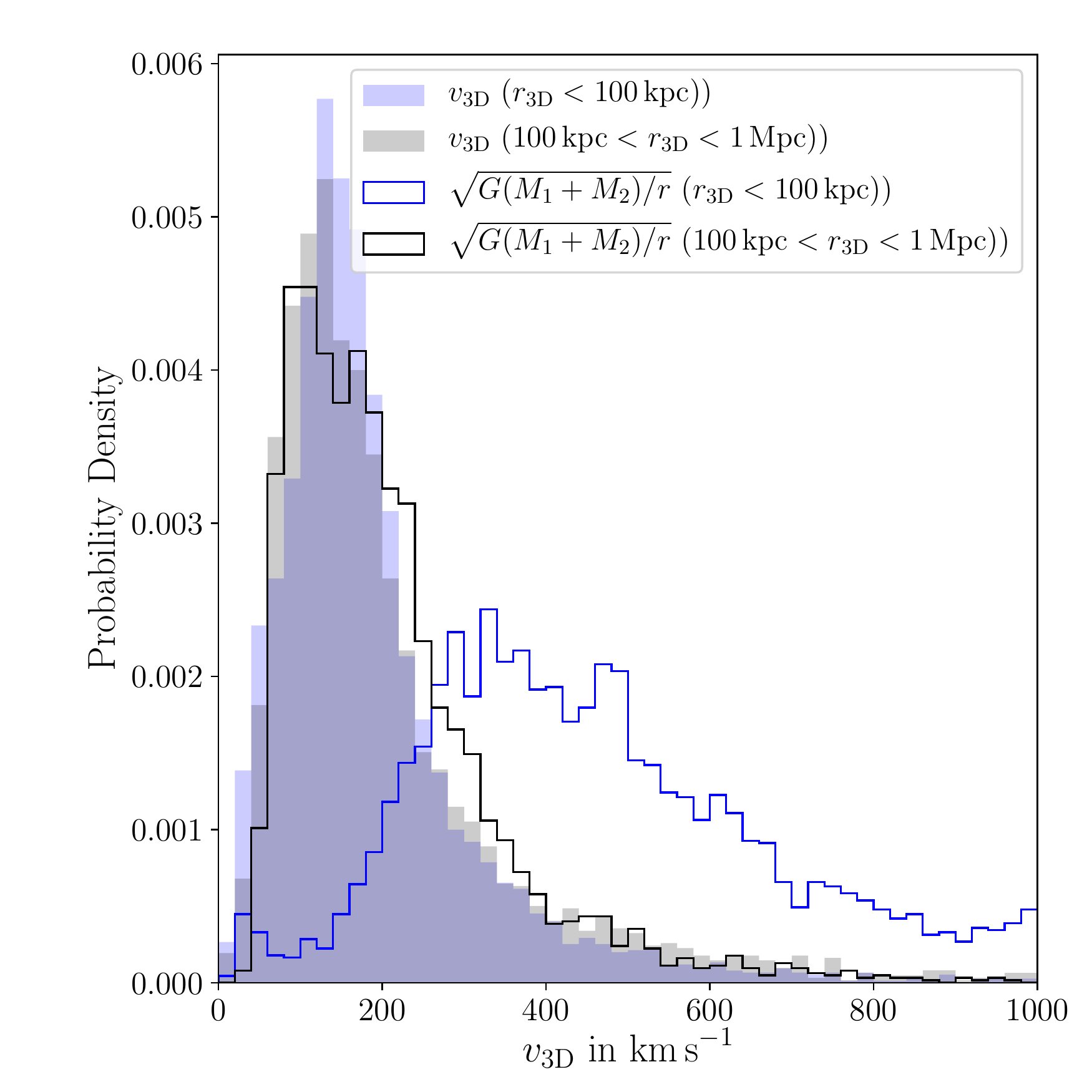}
      \caption{
      Three-dimensional intervelocity $v_\mathrm{3D}$\ and calculated circular velocity $v_\mathrm{circ}$\ for galaxy pairs identified in the simulation with separations $r_\mathrm{3D} < 100\,\mathrm{kpc}$\ (blue) and $100\,\mathrm{kpc}< r_\mathrm{3D} < 1\,\mathrm{Mpc}$. For sufficiently separated pairs, the circular velocity provides a good estimate of the 3D intervelocity, suggesting a physical origin of the intervelocity peak driven by the typical masses and separations of the selected galaxy pairs.
              }
         \label{FigVcirc}
   \end{figure}

We test the latter in our simulation by calculating the circular velocity $v_\mathrm{circ} = \sqrt{G \frac{M_1 + M_2}{r_\mathrm{3D}}}$\ of each galaxy pair, using the total mass of each galaxy, $M_{1,2}$, obtained from the galaxy catalog of the simulation. We ignore pairs with separations exceeding 1\,Mpc as these are most likely non-physical contamination. The distribution of $v_\mathrm{circ}$\ as well as the full 3D velocity difference $v_\mathrm{3D}$\ of the galaxy pairs is plotted in Fig. \ref{FigVcirc}. For pairs with separations of 100\,kpc to 1\,Mpc, $v_\mathrm{circ}$ provides an excellent match to $v_\mathrm{3D}$\ with both showing a pronounced peak at $\sim 130\,\mathrm{km\,s}^{-1}$, and a median $v_\mathrm{3D}$\ in the simulation of 161.9\,km\,s$^{-1}$\ while $v_\mathrm{circ}$\ for the median galaxy masses and separation is 157.5\,km\,s$^{-1}$. However, for pairs with separations closer than 100\,kpc, the circular velocity calculation over-estimates their 3D velocity difference. This is expected, as the velocities of interacting galaxies will be decreased due to dynamical friction once the halos overlap, and the overlapping halos invalidate the implicit assumption of point masses, such that the effective masses in the calculation are smaller than the total masses $M_1$\ and $M_2$.


\section{Discussion and Conclusions}

Pairs of galaxies identified in the observational galaxy catalog HyperLEDA \citep{2014A&A...570A..13M, 2018AstBu..73..310N} have been found to display a peak in their deprojected (three-dimensional) intervelocity distribution at $v_\mathrm{deproj} = 130-150$\,km\,s$^{-1}$\ \citep{2020A&A...641A.115N, 2022MNRAS.510.2167S, 2022arXiv220213766S}. To determine whether such a feature is expected in a cosmological context within $\Lambda$CDM, we have mock-observed galaxies from the IllustrisTNG-300 simulation \citep{2019ComAC...6....2N} and applied the observational approach to identify galaxy pairs. We recover a pronounced peak in the intervelocity of isolated pairs, both via a deprojection of line-of-sight velocity differences \citep{2018A&A...614A..45N} at $v_\mathrm{deproj} = 125\pm7\,\mathrm{km\,s}^{-1}$, and via the direct three-dimensional intervelocity of the identified galaxy pairs at $v_\mathrm{3D} = 132\pm1\,\mathrm{km\,s}^{-1}$.

The agreement between the position of the intervelocity peak of the observational studies and the peak we identified in the simulation is striking, and can be considered a success of the simulation -- and by extension $\Lambda$CDM -- in reproducing this observed feature. It is likely of physical origin since it is not present in randomized galaxy samples, set by the majority of identified pairs for which both galaxies belong to a common dark matter halo, and its position scales with the pair primary's luminosity and thus mass.
It appears plausible that the intervelocity peak of galaxy pairs as a probe of the total mass of galaxies, is related to abundance matching or the stellar mass-halo mass (SMHM) relation. Via these the considered range of galaxy luminosities translates into a corresponding range in total masses, which in turn determine the dynamics leading to a preferred intervelocity. That this intervelocity in $\Lambda$CDM is consistent with the one expected in MOND might then be because the TNG model was devised to repoduce the SMHM relation \citep{2018MNRAS.475..648P}. Since abundance matching such as that of \citet{2013ApJ...770...57B} is consistent with the observed baryonic Tully-Fisher relation (see Fig. 4 of \citealt{2022ApJ...927..198L}), which in turn matches MOND predictions, this might explain why the considered galaxy luminosities in this cosmological simulation are tied to just the right total halo mass to reproduce the observed feature of the intervelocity distribution.

Recovering the intervelocity peak with a $\Lambda$CDM simulation does not reduce MOND's success in reproducing it, but it shows that (1) MOND is not the only model offering an explanation, and (2) $\Lambda$CDM is not immediately challenged by the observed intervelocity peak of galaxy pairs. This does not mean that the dynamics of isolated galaxy pairs can not be developed into a useful tool to test gravitational dynamics in the low-acceleration regime. However, this will require more dedicated research to determine how the peak comes about in the cosmological context, in what details predictions of the intervelocity distributions between $\Lambda$CDM, MOND, and other models differ (e.g. are there environmental effects), and what observational data will be required to discriminate these alternatives.


\begin{acknowledgements}
We thank Mordehai Milgrom for interesting discussions, and the anonymous referee for helpful suggestions, that improved our manuscript. MSP acknowledges funding of a Leibniz-Junior Research Group (project number J94/2020) and a KT Boost Fund by the German Scholars Organization and Klaus Tschira Stiftung. PL is supported by the Alexander von Humboldt Foundation. 
\end{acknowledgements}

%
   \bibliographystyle{aa} 
   \bibliography{GalaxyPairs} 
%


\begin{appendix} 
\section{Comparisons of the Observed and Simulated Galaxy Catalogs}
\label{appendix:othertests}

   \begin{figure*}
   \centering
   \includegraphics[width=0.33\hsize]{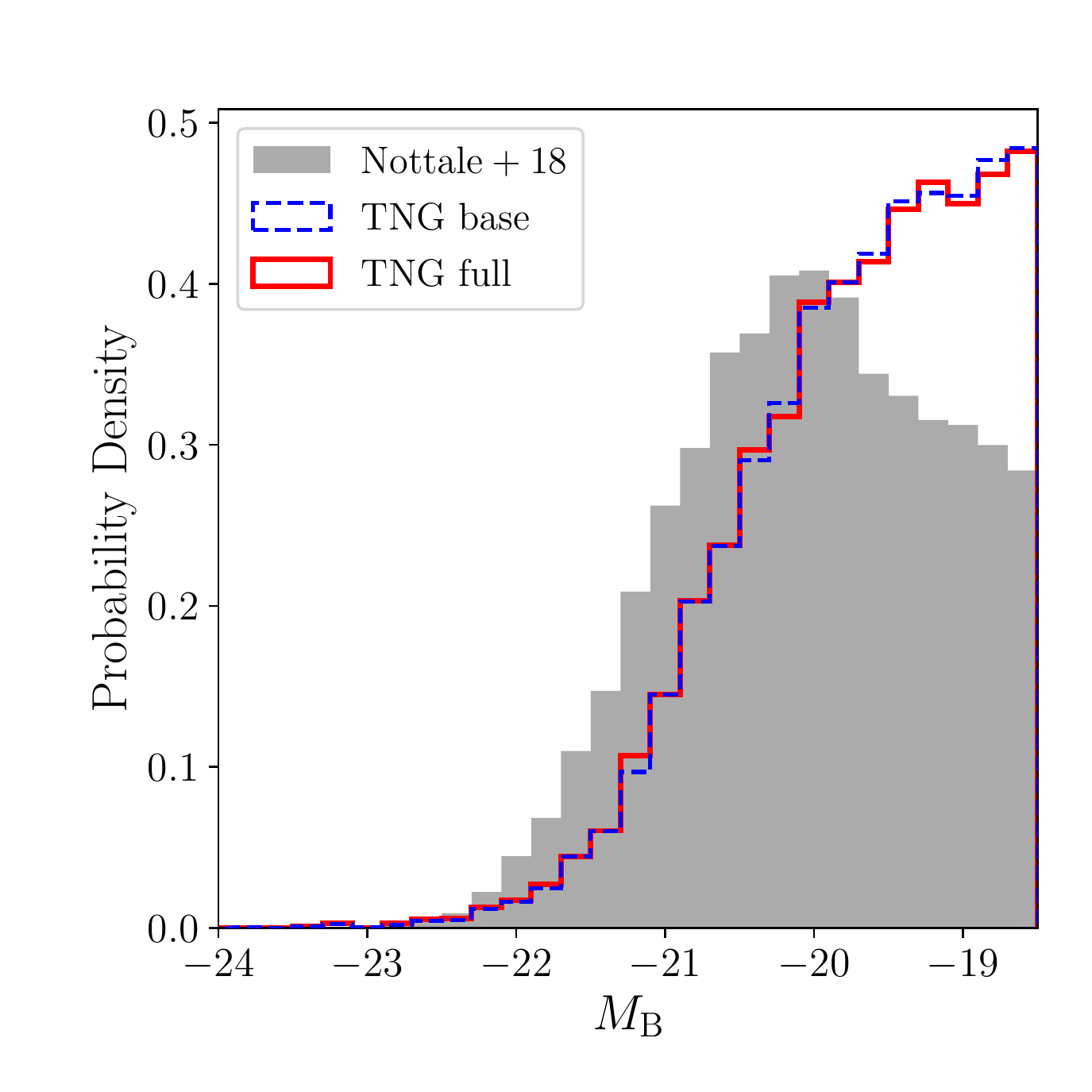}
   \includegraphics[width=0.33\hsize]{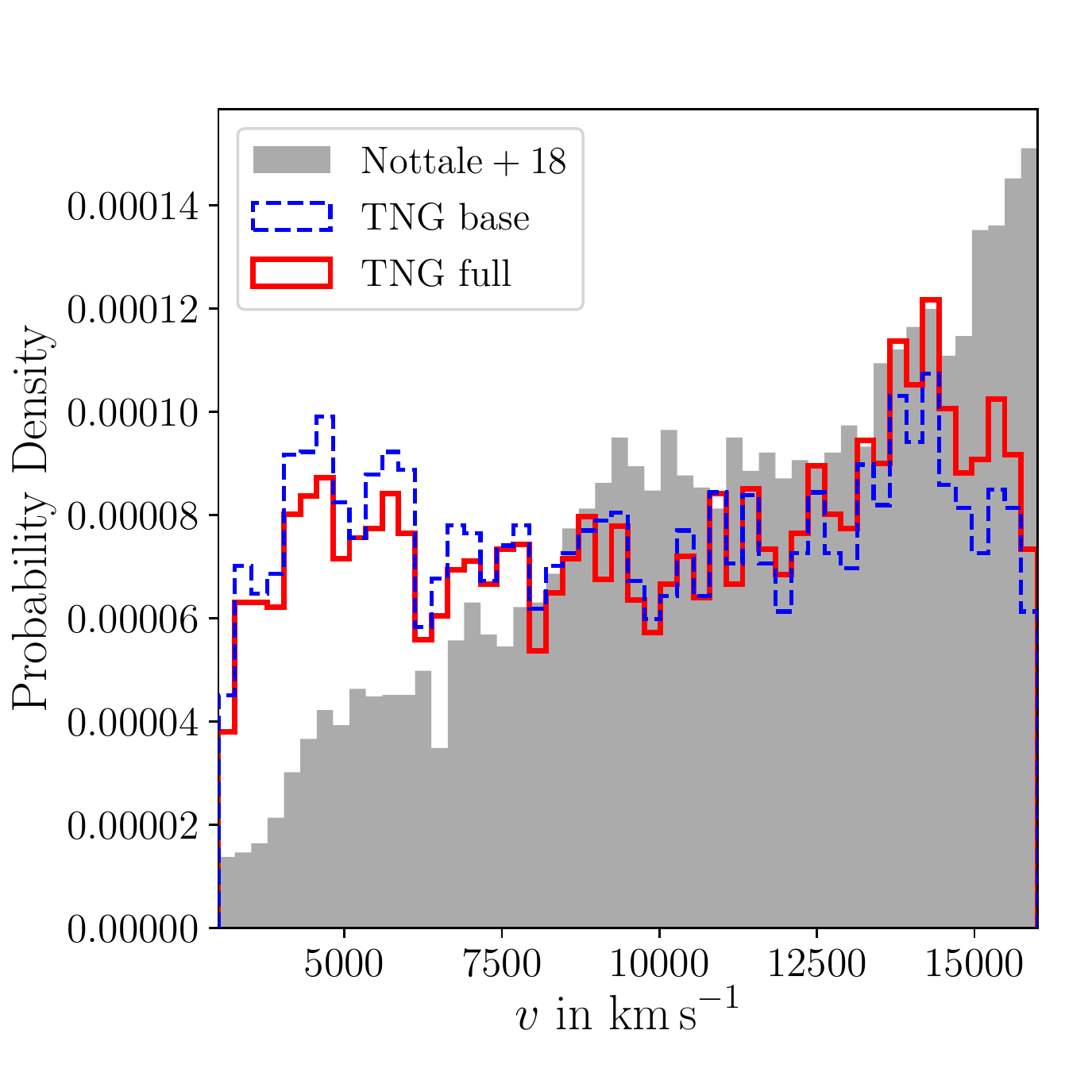}
      \caption{Properties of galaxy pairs for the observed  galaxy pair catalog properties from \citet{2018AstBu..73..310N} (grey shaded histograms), and galaxy data from the IlustrisTNG-300 simulation. For the latter, the blue histogram corresponds to our base model without velocity errors, and the red histograms are from our full model mimicking apparent magnitude biases and errors on the observed line-of-sight velocity.
      {\it Left panel}: Distribution of galaxy pair members in B-band magnitude $M_\mathrm{B}$.
      {\it Right panel}: Distribution of galaxy pairs in velocity $v$, consisting of redshift and peculiar velocity. 
              }
         \label{FigSample1}
   \end{figure*}

   \begin{figure*}
   \centering
   \includegraphics[width=0.33\hsize]{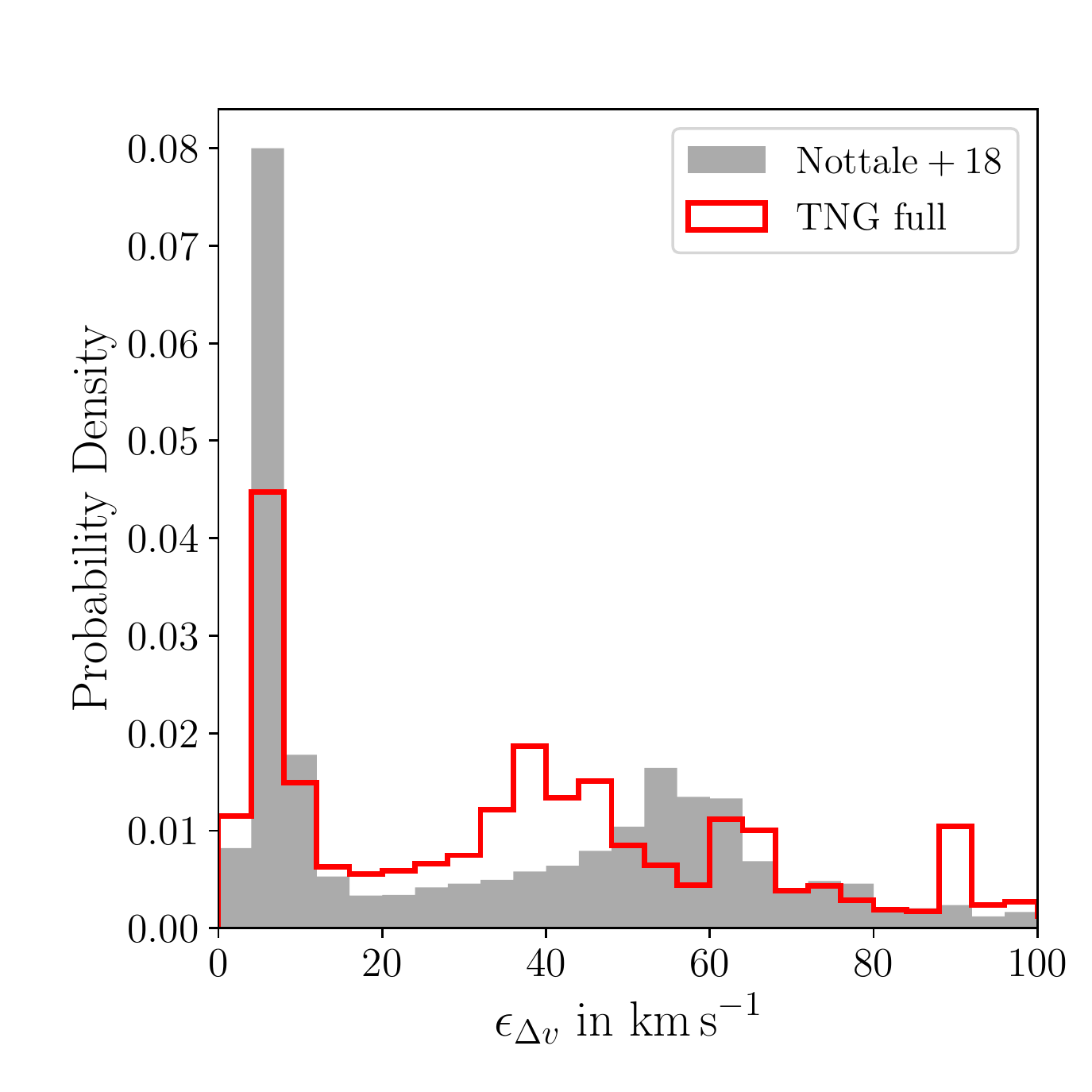}
   \includegraphics[width=0.33\hsize]{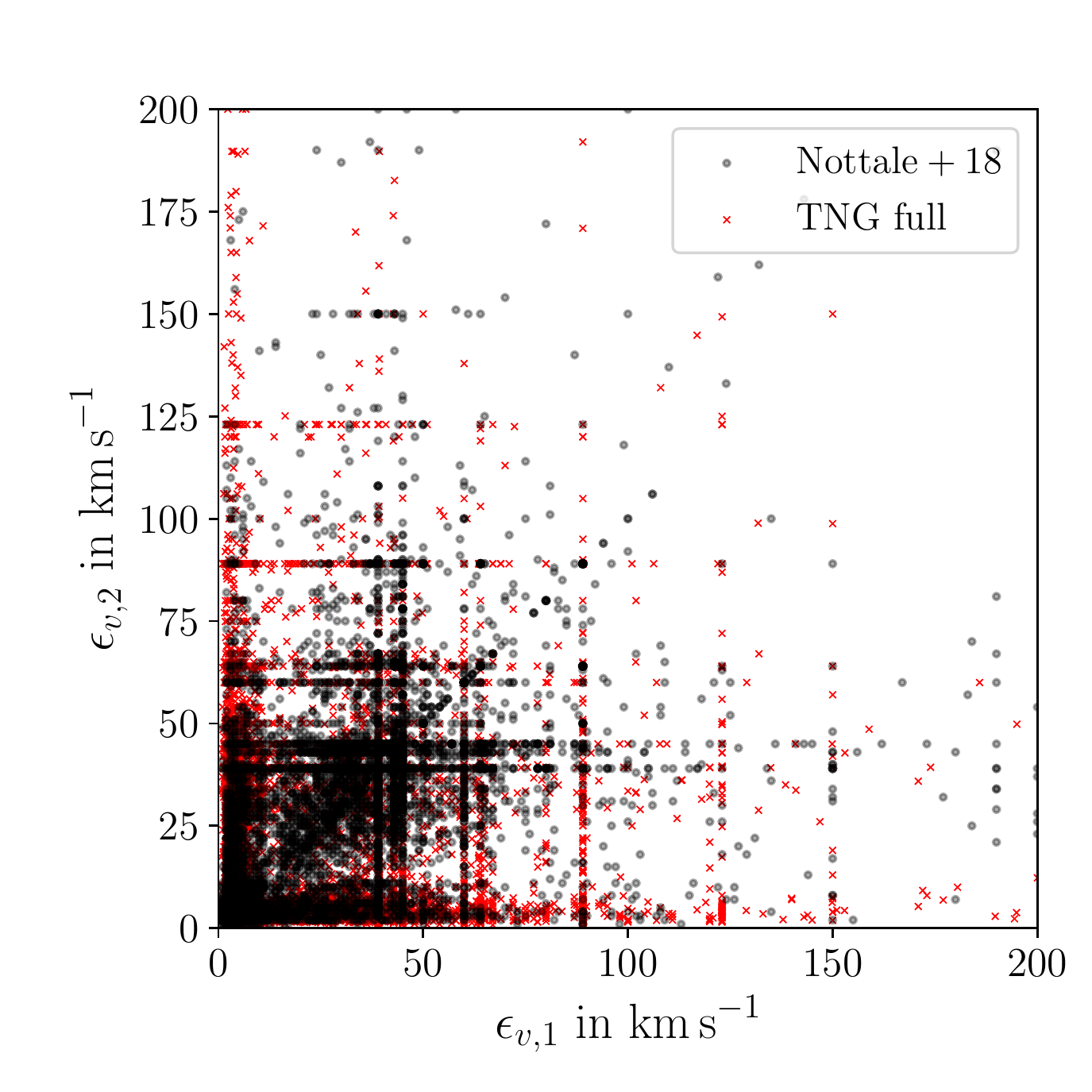}
      \caption{Comparison of velocity errors for the observed and simulated galaxy pair samples.
      {\it Left panel}: Distribution of errors $\epsilon_{\Delta v}$\ on the line-of-sight velocity difference $\Delta v$\ of galaxy pairs.
      {\it Right panel}: Errors on individual velocity measurements for the two corresponding partner galaxies (more luminous partner on the horizontal axis). Some error correlation is present for the observed sample, but absent in our mock observed simulation data.
              }
         \label{FigSample2}
   \end{figure*}

While our mock-observational approach and our pair finding algorithm closely follow that of the observational study, a number of differences are present: (1) our simplified approach does not cover the same volume in shape as the observational galaxy catalog, (2) the simulation is not affected by spatial biases, obscuration, or crowding issues, and (3) we employ a simplified treatment of velocity measurement errors.

The distribution of galaxy pairs in absolute B-band magnitude $M_\mathrm{B}$ and redshift velocity $v$\ is shown in Fig. \ref{FigSample1}. The former shows the similarity in the covered magnitude range, though the observational sample is lacking in galaxies fainter than $M_\mathrm{B} = -20$. This is likely due to insufficient completeness of the observational catalog as compared to the complete galaxy catalog available for the simulation. 
The distributions in velocity highlight the difference in the selection volumes of the observational study and our mock-observational approach using the IllustrisTNG-300 simulations. The observational sample rises with redshift, as a larger volume of spherical shells contributes to the galaxy distribution. Such a volume is not feasible due to the limited simulation box size. Consequently, the apparent magnitude bias resulting in less rejected pairs at higher redshifts discussed in \citet{2018AstBu..73..310N} is not as severe for our mock-observed simulation samples. We can therefore expect the latter to be less contaminated by non-isolated pairs.

Fig. \ref{FigSample2} compares the error of the pair velocity difference, $\epsilon_{\Delta v}$, for the observational pair catalog and the one obtained for our full sample, for which velocity errors are drawn from the errors of the HyperLEDA galaxy database. This approach typically results in similar errors for pairs: a strong and sharp peak at small $\epsilon_{\Delta v} < 10\,\mathrm{km\,s}^{-1}$, and another broader peak at higher $\epsilon_{\Delta v}$. However, there are some differences in the details, such as the relative height of the first peak and the position of the second one. We suspect that some correlation between galaxy errors in the observational data is a possible cause of this.

This appears to be confirmed by the right panel of Fig. \ref{FigSample2}, which plots the individual velocity errors for the two members of a pair against each other. Some of the observed pairs indeed show correlation between their velocity errors. This is not unexpected, as galaxies close to each other on the sky might have likely been measured with the same technique, instrument, or survey, resulting in them having similar errors. This can result in low-error galaxies as well as high-error galaxies being predominantly associated with each other, compared to a completely random error assignment. Such effects are difficult to accurately reproduce in mock-observing the simulation. However, they are not expected to significantly alter the results, because the velocities are only relevant for the pair finding algorithm -- where they are generally smaller than the allowed velocity difference of 500\,km\,s$^{-1}$\ -- and for the deprojected intervelocity -- which however turns out to agree with the error-free, three-dimensional intervelocity peak of galaxy pairs in the simulation. This is confirmed by the absence of substantial differences between the results for our base and full input catalogs, despite the former being entirely devoid of any velocity errors.

\section{Deprojected Velocities for Projections Along Other Axes}
\label{Appendix:DifferentDirections}

   \begin{figure*}
   \centering
   \includegraphics[width=0.3\hsize]{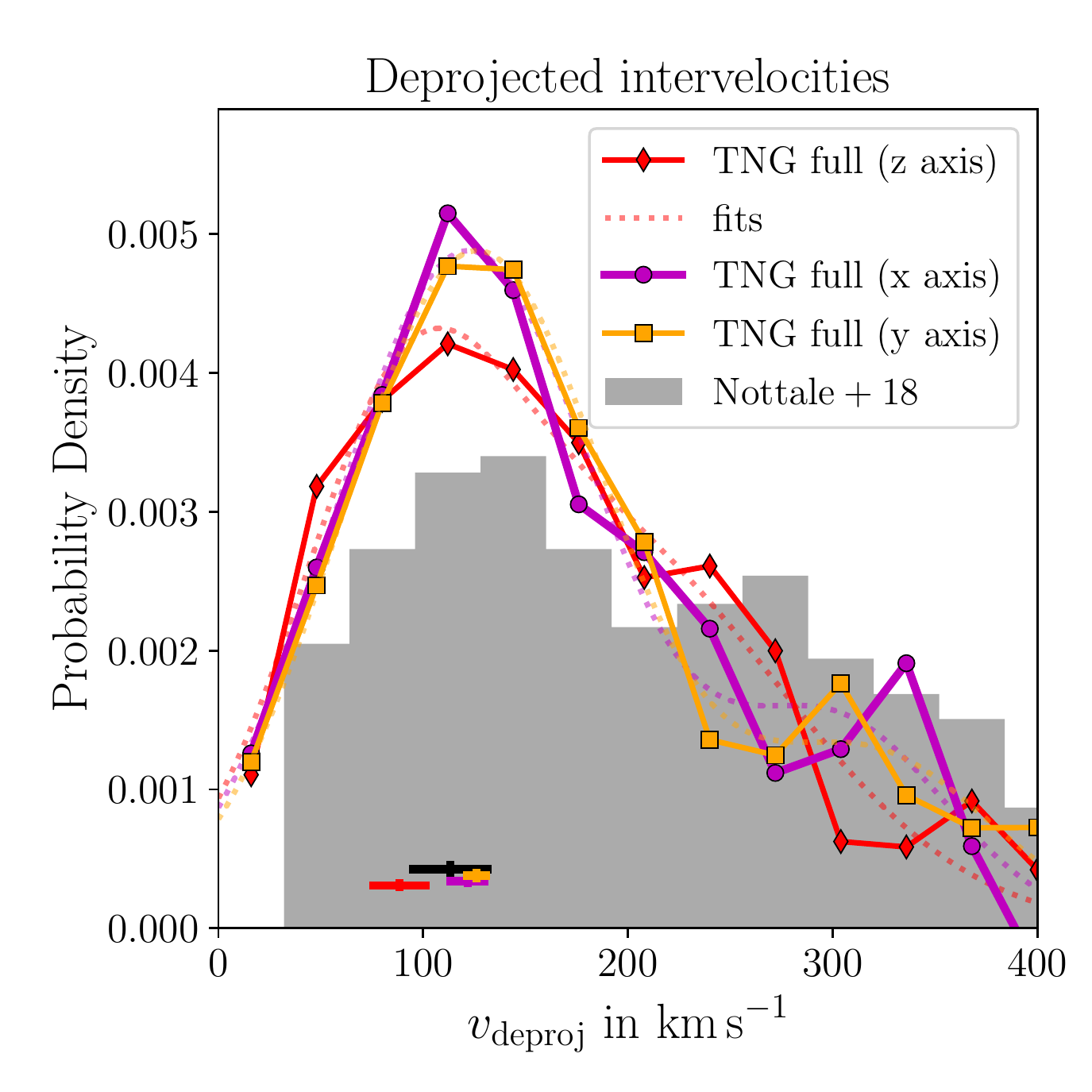}
   \includegraphics[width=0.3\hsize]{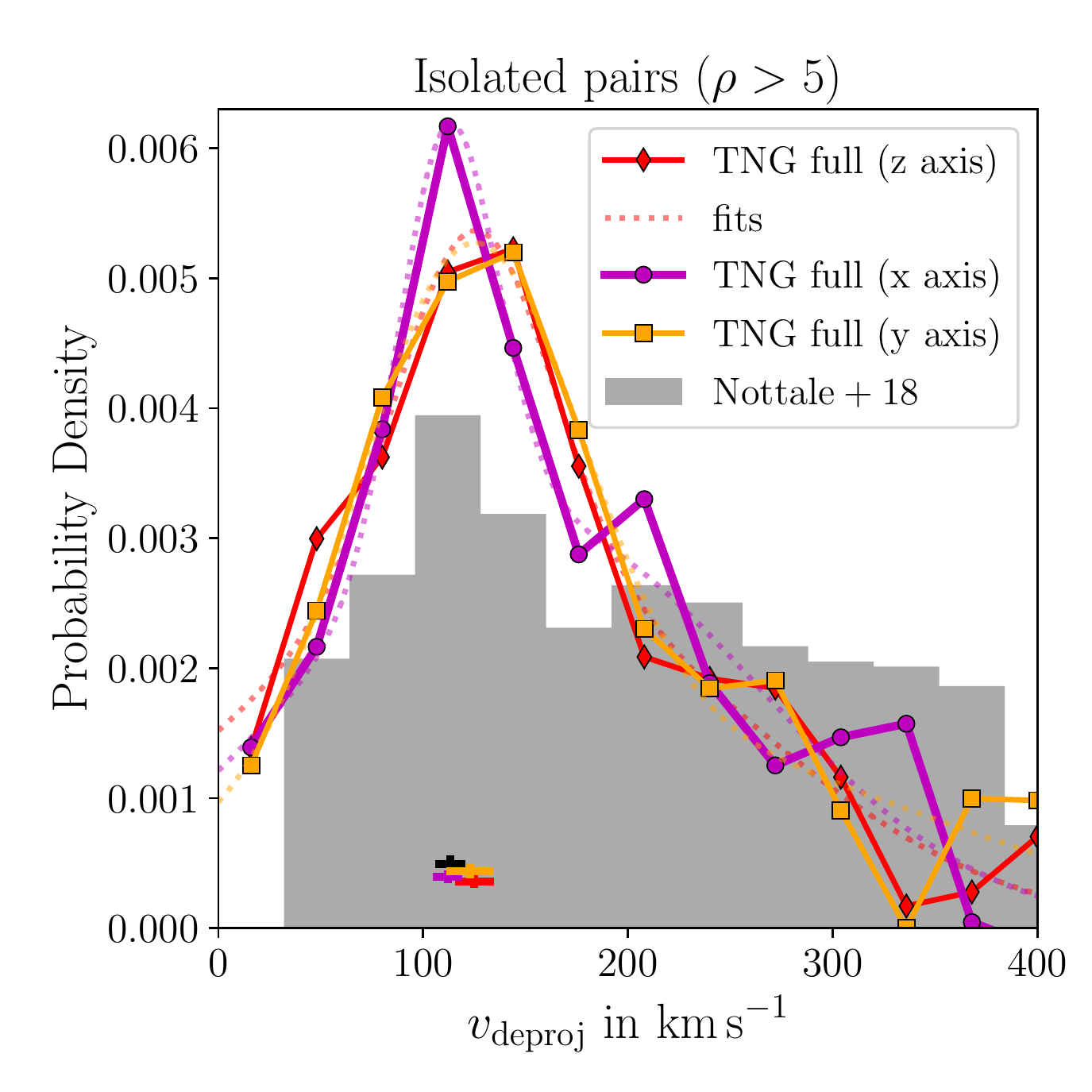}
   \includegraphics[width=0.3\hsize]{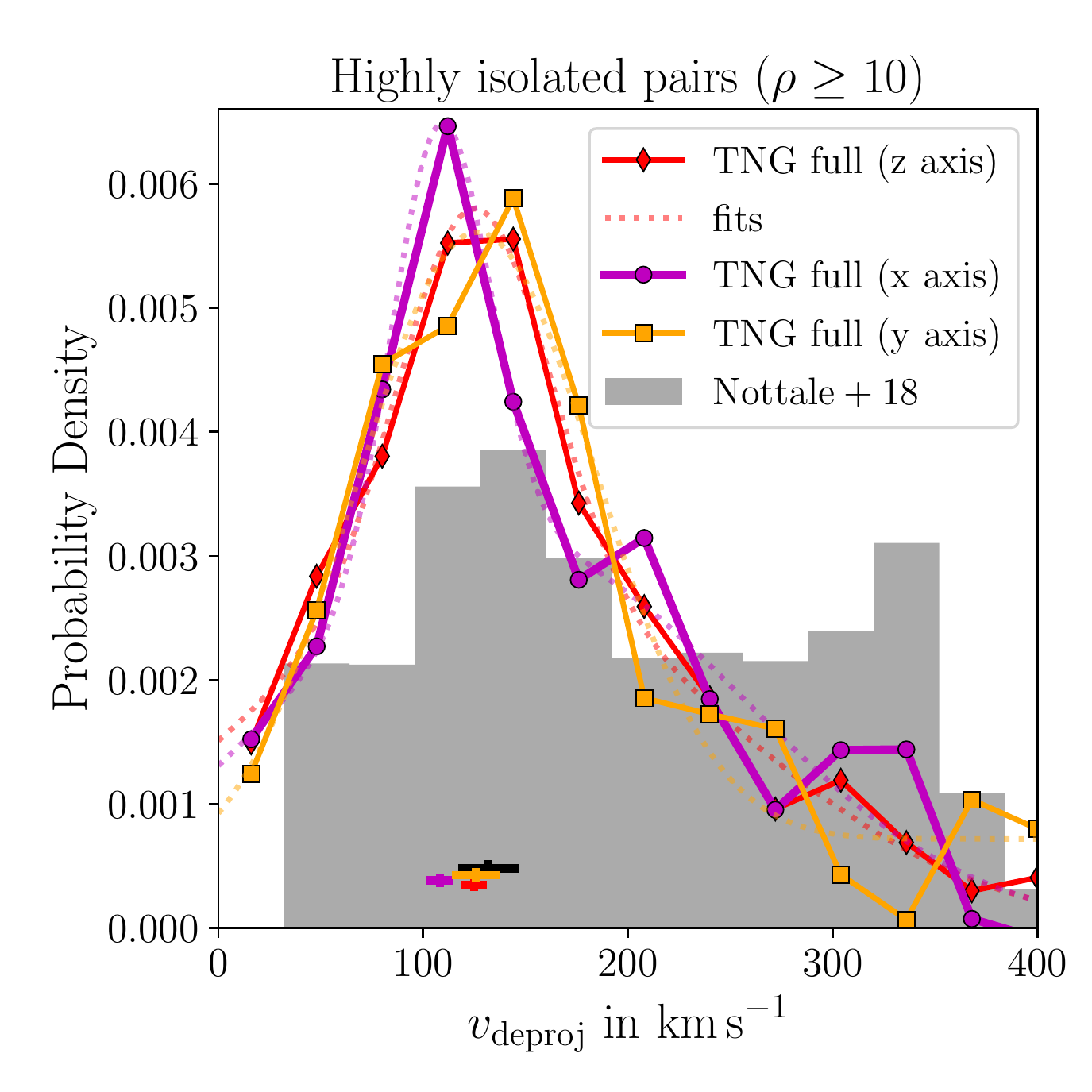}
      \caption{Deprojected velocities, similar to Fig. \ref{FigDeprojections} but showing the mock-observed simulation along the three different axes of the simulation box.
             }
         \label{FigDeprojectionsAxes}
   \end{figure*}

In Fig. \ref{FigDeprojectionsAxes} we plot the intervelocities of galaxy pairs obtained from the deprojection approach of Sect. \ref{Sect:Intervelocity} for the mock-observed (full) galaxy samples projected along the three different axes of the simulation box. Overall, the presence of the intervelocity peak is stable and its peak position is consistent with that of the observed galaxy pairs. However, some differences in the deprojected intervelocity distributions along the three axes are present and illustrate the degree of uncertainty in the velocity deprojection approach.

   \begin{figure}
   \centering
   \includegraphics[width=\hsize]{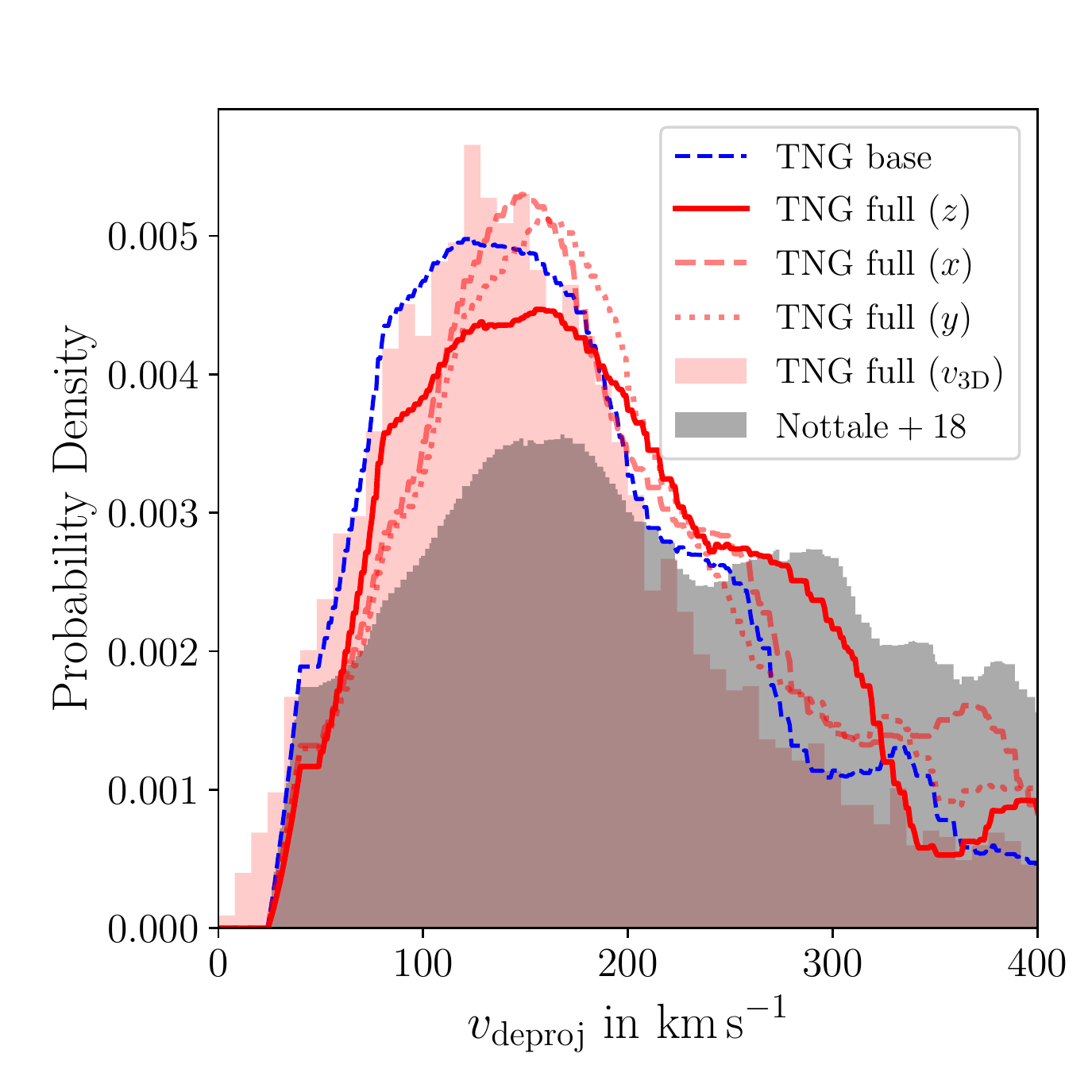}
      \caption{Deprojected intervelocities obtained from averaging deprojections with bin width between 25 and 40\,km\,s$^{-1}$, following the approach introduced by \citet{2022arXiv220213766S}.
         }
         \label{FigMultibinaverage}
   \end{figure}

To more accurately determine the position of the deprojected intervelocity peak despite the wide velocity bins required by the deprojection algorithm, \citet{2022arXiv220213766S} have performed the deprojection with a range of bin width and then averaged the resulting histograms by re-sampling with a bin width of 1\,km\,s$^{-1}$. For completeness, we follow their approach for our simulation data in Fig. \ref{FigMultibinaverage}. Specifically, we apply the deprojection algorithm for bin width between 25 and 40\,km\,s$^{-1}$, re-sample the resulting de-projected histograms in bins of 1\,km\,s$^{-1}$\ width, and average them. This results in a smoother deprojected velocity distribution that helps to identify the position of the intervelocity peak, though we warn that the choice of bin width and range, as well as numerical noise due to relatively small pair sample sizes, can introduce difficult to quantify biases and inaccuracies in this method. This is illustrated by showing not only the $z$-axis projection for the simulations, but also those for the $x$\ and $y$ directions. The different directions of projection of the simulated galaxy distribution also result in some differences in the re-sampled deprojected intervelocity curves, the amount of which gives an idea of the level to which such curves can be trusted.

\end{appendix}

\end{document}